\documentclass[12pt]{iopart}

\usepackage{iopams}  
\usepackage{graphicx}
\usepackage{color}
\usepackage{ulem}
\newcommand{\rev}[1]{#1}
\newcommand{\del}[1]{}

\begin{document}

\title[Morse index for figure-eight choreographies]
{Morse index for figure-eight choreographies of the \rev{planar} equal mass three-body problem}

\author{Hiroshi Fukuda$^1$, Toshiaki Fujiwara$^1$ and Hiroshi Ozaki$^2$}

\address{$^1$ College of Liberal Arts and Sciences, Kitasato University, 1-15-1 Kitasato, Sagamihara, Kanagawa 252-0329, Japan}
\address{$^2$ Laboratory of general education for science and technology, Faculty of Science, Tokai University, 
4-1-1 Kita-Kaname, Hiratsuka, Kanagawa, 259-1292, Japan}
\ead{fukuda@kitasato-u.ac.jp, fujiwara@kitasato-u.ac.jp and ozaki@tokai-u.jp}
\vspace{10 pt}
\begin{indented}
\item[] \today 
\end{indented}

\begin{abstract}
We report on numerical calculations of Morse index for figure-eight choreographic solutions to 
a system of three identical bodies 
\rev{in a plane}
interacting through homogeneous potential, $-1/r^a$, or 
through Lennard-Jones-type (LJ) potential, $1/r^{12} - 1/r^6$,
where $r$ is a distance between the bodies. 
The Morse index is a number of independent variational functions 
giving negative second variation $S^{(2)}$ of action functional $S$.
We calculated three kinds of Morse indices, $N$, $N_c$ and $N_e$, 
in the domain of the periodic, the choreographic and 
the figure-eight choreographic function, respectively.
For homogeneous system, we obtain 
$N=4$ for $0 \le a < a_0$, 
$N=2$ for $a_0 < a < a_1$,  
$N=0$ for $a_1 < a$, and
$N_c=N_e=0$ for $0 \le a$, 
where
$a_0=0.9966$ 
and $a_1=1.3424$. 
For $a=1$, we show a strong relationship between the figure-eight choreography 
and the periodic solution found by Sim\'{o} through the $S^{(2)}$.  
For LJ system, we calculated the index 
for the solution tending to the figure-eight solution of $a=6$ homogeneous system 
for the period $T \to \infty$.
We obtain $N$, $N_c$ and $N_e$ 
as monotonically increasing functions of the gradual change in $T$ from $T \to \infty$,
which start with $N=N_c=N_e=0$, 
jump at the smallest $T$ 
by $1$, and
reach $N=12$, $N_c=4$, and $N_e=1$ for $T \to \infty$ in the other branch.
\end{abstract}

%
%
\submitto{\JPA}
%
%
%

\section{Introduction}
\label{sec:intro}
Choreographic motion of $N$ bodies is a periodic motion on a closed orbit, 
$N$ identical bodies chase each other on the orbit with equal time-spacing. 
Moore \cite{moore} found a remarkable figure-eight three-body choreographic solution 
under homogeneous potential $-1/r^a$ by numerical calculations, 
where $r$ is a distance between bodies. 
Chenciner and Montgomery \cite{chenAndMont} gave a mathematical proof 
of its existence for $a=1$
by variational method. 
\rev{The detailed initial conditions for three bodies are found in \cite{chenAndMont,simoH}.}

Sbano \cite{sbano2005}, Sbano and Southall \cite{sbano}, and Fukuda {\it et~al} \cite{fukuda},
after that, 
studied $N$-body choreographic solutions 
under an inhomogeneous potential 
\begin{equation}
   u^{LJ}(r)=\frac{1}{r^{12}}-\frac{1}{r^6},
\label{eq:LJu}
\end{equation}
a model potential between atoms called Lennard-Jones-type (hereafter LJ) potential.
Sbano and Southall \cite{sbano} proved that 
there exist at least two $N$-body choreographic solutions for sufficiently large period $T$, 
and there exists no solution for small period $T$. 
Then we confirmed their theorem numerically and unexpectedly
found a multitude of figure-eight choreographic solutions
under LJ potential (\ref{eq:LJu})  \cite{fukuda}.

Recently, Shibayama \cite{shibayama}
calculated Morse index numerically for 
the figure-eight and for super-eight choreography 
to consider 
the variational proof of their existence. 
Here Morse index is a number of independent variational functions giving negative second variation
of action functional. 

There are several researches on Morse indices for periodic solution of three body problem.
Barutello {\it et~al} \cite{barutello} calculated Morse index mathematically for the Lagrangian circular orbit,
and Hu and Sun \cite{hu,hu2} for elliptic Lagrangian solutions,  
to discuss the linear stability.

In this paper, we calculate Morse indices numerically
for the figure-eight choreographies to a system of three identical bodies 
interacting through a homogeneous potential 
or through LJ potential (\ref{eq:LJu}). 
We expect that accurate numerical calculations of Morse index will reveal 
their structures and relations via the geometry of their action manifolds.
In section \ref{sec:morse}, 
we define Morse index 
and present corresponding eigenvalue problem and our method of its numerical calculation.
In section \ref{sec:homo},
Morse index for the system interacting through homogeneous potential 
with various $a \ge 0$ are calculated.
For $a=1$ we point out strong relationship between the figure-eight choreography and 
periodic solution close to it found by Sim\'{o} \cite{simoH} through the 
second variation
of action functional.
In section \ref{sec:LJ},
we calculate the Morse index for a solution we found \cite{fukuda}
in the system interacting through LJ potential (\ref{eq:LJu}),
tending toward the figure-eight choreography in the homogeneous system with $a=6$
for $T \to \infty$.
We discuss the Euler 
\rev{characteristic} of their action manifold. 
Further the correspondence of the results between LJ and homogeneous system is investigated. 
Section \ref{sec:summary} is a summary and discussions.
Our numerical results in this paper were calculated by Mathematica 11.1 in its default precision, 
unless otherwise stated.

\section{Numerical calculation of Morse index}
\label{sec:morse}

\subsection{Eigenvalue problem for Morse index}
\label{sec:eigen}
For a system of three identical bodies in classical mechanics, 
we consider periodic solutions to equations of motion, 
\begin{equation}\label{eq:motion}
  \frac{d}{d t} \frac{\partial L}{\partial \dot{q}_i} = \frac{\partial L}{\partial q_i},
	\rev{\; i=1,2, \ldots, 6,}
\end{equation}
where dot represents a differentiation in $t$.
$L$ is the Lagrangian with the potential energy $U(q)$
\begin{equation}\label{eq:L}
  L(q,\dot{q})
  =\sum_{i=\rev{1}}^{\rev{6}}\frac{\dot{q}_i^2}{2}-U(q),   
\end{equation}
and 
\begin{equation}\label{key}
	q(t)=(q_{\rev{1}}(t), q_{\rev{2}}(t),\ldots, q_{\rev{6}}(t))^*
\end{equation}
a six component vector composed of position vectors 
\begin{equation}\label{key}
	\rev{\bi{r}_b}(t) = (x_{\rev{b}}(t), y_{\rev{b}}(t))^* = (q_{\rev{2 b-1}}(t), q_{\rev{2 b}}(t))^*
\end{equation}
for body \rev{$b=1,2,3$} 
moving in a plane, 
where $^*$ represents transpose.

For a periodic solution $q(t+T)=q(t)$ with period $T$,
we calculate the second variation of the action
\begin{equation}\label{key}
  S(q)=\int_{0}^{T} L(q,\dot{q}) d t.
\end{equation}
The $k$'th variation $S^{(k)}$ of the action $S(q)$ is defined as the $k$'th coefficients in
\begin{equation}\label{eq:S}
  S(q+h \delta q)=S^{(0)}+h S^{(1)}+\frac{h^2}{2!} S^{(2)}+\cdots,
\end{equation}
thus
\begin{equation}\label{key}
  S^{(k)}=\int_{0}^{T}d t \left( \sum_{i}(\delta q_i \frac{\partial}{\partial q_i}+
  \dot{\delta q}_i \frac{\partial}{\partial \dot{q}_i}) \right) ^k L,
\end{equation}
where $h$ is a real number and $\delta q$ is a variation function with period $T$,
$\delta q(t+T)=\delta q(t)$.

By partial integration, the second variation is written as
\begin{equation}\label{key}
  S^{(2)}=(\delta q, \hat{H} \delta q)
\end{equation}
by $6\times 6$ matrix operator $\hat{H}$, 
\begin{equation}\label{key}
  \hat{H}_{i j} = -\delta_{i j} \frac{d^2}{d t^2} - U_{i j}(t), 
\end{equation}
with 
\begin{equation}\label{eq:Umat}
  U_{i j}(t) 
  = \frac{\partial^2 U}{\partial q_i \partial q_j}.
\end{equation}
The inner product $(f,g)$ is defined as 
\begin{equation}\label{key}
  (f,g)=\int_{0}^{T} d t f^* g
\end{equation}
and $\delta_{i j}$ the Kronecker delta.
Considering eigenvalue $\lambda$ and eigenfunction $\psi$   
of the operator $\hat{H}$, 
\begin{equation}\label{eq:eigen}
  \hat{H} \psi = \lambda \psi,
\end{equation}
the second variation for $\delta q=\psi$ is given by 
\begin{equation}\label{key}
  S^{(2)}=\lambda.
\end{equation}
Then the Morse index is the number of negative eigenvalues of (\ref{eq:eigen}). 
Here the eigenfunction $\psi$ is assumed to be normalized as 
\begin{equation}\label{eq:norm}
  (\psi,\psi)=1. 
\end{equation}

\subsection{Figure-eight choreographic, choreographic and non-choreographic eigenfunction}
\label{sec:choreo}
We consider the eigenvalue problem (\ref{eq:eigen}) for a figure-eight choreography $q$.
A function $f$ is called \rev{\textit{choreography}} or \rev{\textit{choreographic}} 
if $f$ satisfies
\begin{equation}\label{eq:Cq}
   \hat{C} f = f
\end{equation}
where the linear operator $\hat{C}$ is defined by
\begin{equation}\label{eq:Chat}
   \hat{C} f_i(t) = f_{i+2}(t-\frac{T}{3}). 
\end{equation}
A \rev{\textit{figure-eight choreography}} is a choreography with its orbit symmetric in $x$- and $y$-axis. 
Here in (\ref{eq:Chat}) and hereafter 
the subscript of six component vector 
\rev{is assumed to be in the range between 1 and 6 with translation by 6.}

Then the eigenfunction $\psi$ with period $T$ of (\ref{eq:eigen}) is classified into the following three types:
%
1) Choreographic eigenfunction if $\psi$ is choreographic, 
which is possible since $\hat{C}$ and $\hat{H}$ commute. 
2) Figure-eight choreographic eigenfunction if $q+h\psi$
is figure-eight choreographic. 
%
3) Non-choreographic eigenfunction is 
a orthogonal complement of choreographic eigenfunction.

Accordingly, we obtain three kind of Morse index at the figure-eight choreography $q$
in different domain from the common eigenvalue problem (\ref{eq:eigen}) for periodic $\psi$: 
Morse index in the domain of the periodic function,
the choreographic function, and 
the figure-eight choreographic function.
In the following we denote these three Morse indices 
in the different domains as 
$N$, $N_c$ and $N_e$, respectively.  

Note that 
variational functions representing 
translation in $x$- and $y$-direction, rotation,
and translation in time, 
keep the action integral $S(q)$ constant, 
and their derivatives are the eigenfunctions of zero eigenvalues.
Therefore the zero eigenvalues of (\ref{eq:eigen}) are quadruply degenerated and 
their eigenfunctions correspond to the conservation law of linear and angular momentum, 
and energy, respectively. 

Further 
the equation (\ref{eq:eigen})
has trivial solutions
\begin{equation}\label{eq:trivial}
  \lambda=k^2\omega^2, \; k=1,2,3,\ldots,
\end{equation}
\begin{equation}\label{eq:trivialfunc}
  \psi(t) =(x(t), y(t), x(t), y(t), x(t), y(t))^* 
\end{equation}
with $x(t)=\sin{k\omega t}$ or $\cos{k\omega t}$, 
and $y(t)=\sin{k\omega t}$ or $\cos{k\omega t}$,
where $\omega=2\pi/T$. 
Since $U(q)$ is functions of 
\rev{$|\bi{r}_{b}-\bi{r}_{c}|$} 
unaffected 
by the variation $\delta q(t)=\psi(t)$ in (\ref{eq:trivialfunc}),
$\delta^2 U =0$, $\sum_j U_{i j} \psi_j =0$ and  $\hat{H} \psi = -d^2\psi/d t^2$.
Thus the solutions (\ref{eq:trivial}) and (\ref{eq:trivialfunc}) are derived.
We call this quadruply degenerated eigenfunctions \rev{\textit{trivial}} \cite{fujiwara}.


\subsection{Fourier series expansion}
\label{sec:fourier}
Following Shibayama \cite{shibayama},
we solve the eigenvalue problem (\ref{eq:eigen}) by expanding the $\psi$ in
the Fourier series
\begin{equation}\label{eq:fourier}
  \psi_i(t)=\sum_{k=0}^{M-1} \rev{v_k^{(i)}}\phi_k(t) 
\end{equation}
where
\begin{equation}\label{key}
  \phi_k(t) =\left\lbrace \begin{array}{l l} 
      \displaystyle \sqrt{\frac{2}{T}}\sin(\frac{k+1}{2} \omega t) & k=1,3,5,\ldots \\
      \displaystyle \sqrt{\frac{2}{T(1+\delta_{k0})}}\cos(\frac{k}{2} \omega t) & k=0,2,4,\ldots \\
    \end{array}\right. 
\end{equation}
are the normalized basis as 
\begin{equation}\label{key}
  \int_{0}^{T} \phi_k \phi_l d t = \delta_{k l}.
\end{equation}
Thus (\ref{eq:eigen}) becomes the eigenvalue problem  
\begin{equation}\label{eq:eigenBmat}
  H v = \lambda v 
\end{equation}
\rev{for}
$6 M \times 6 M$ real symmetric matrix $H$
\begin{equation}\label{eq:bmat}
\rev{H_{6k+i,6l+j}}
  =\int_0^T d t \phi_k \hat{H}_{i j} \phi_l 
  = 
\rev{-u_{6k+i,6l+j}}
  +  \omega^2 \lfloor \frac{k+1}{2} \rfloor^2 \delta_{i j}\delta_{k l},
\end{equation}
where
\begin{equation}\label{eq:umat}
\rev{u_{6k+i,6l+j}}
	=\int_{0}^{T} d t \phi_k U_{i j}(t) \phi_l
\end{equation}
and $\lfloor.\rfloor$ is the floor function. 
The vector $v$ is a column vector of $6 M$ components, 
$\rev{v_{6 k+i}=v_k^{(i)}, }$
with 
\begin{equation}\label{key}
  v^* v=1 
\end{equation}
by the normalization condition (\ref{eq:norm}).

The matrix elements (\ref{eq:umat}) are calculated from about $42 M$ integrals 
\begin{equation}\label{eq:ukmat}
  u_{i j}(k)=u_{i j}^{(+)}(k)+\imath u_{i j}^{(-)}(k)
  =\frac{2}{T}\int_{0}^{T} d t \exp(\imath k\omega t) U_{i j}(t)
\end{equation}
as
\begin{equation}\label{eq:ukmat2}
\rev{u_{6k+i,6l+j}}
  =\left\lbrace \begin{array}{l  l}
  \displaystyle 
  \frac{p^l}{2} u_{i j}^{(p)}(k'-l')
  + \frac{(-1)^{k l}}{2} u_{i j}^{(p)}(k'+l'), & (k>0, l>0)\\
  \displaystyle 
  \frac{1}{\sqrt{1+\delta_{k 0}}\sqrt{1+\delta_{l 0}}}u_{i j}^{(p)}(k'+l'), & (\mbox{otherwise})
  \end{array}\right. 
\end{equation}
where
\begin{equation}\label{key}
  p=(-1)^{k+l}, \; k'=\lfloor\frac{k+1}{2}\rfloor, \; l'=\lfloor\frac{l+1}{2}\rfloor,
\end{equation}
and $\imath=\sqrt{-1}$.
Though the upper first term in (\ref{eq:ukmat2}) looks non symmetric in $k$ and $l$
at a glance, it is symmetric as defined by (\ref{eq:umat})
since $p^l=-p^k$ 
for $p=-1$.

We evaluate the integral (\ref{eq:ukmat}) with periodic integrand efficiently 
by trapezoidal formula of numerical integration 
with $n$ points and it is done by fast Fourier transform quickly.

\section{Homogeneous potential}
\label{sec:homo}
For 
the system 
interacting through the homogeneous potential
\begin{equation}\label{eq:homou}
  U(q)= 
  -\sum_{\rev{b>c}}\frac{1}{r_{\rev{b c}}^a}
\end{equation}
where
\rev{$  r_{b c}=|\bi{r}_{b}-\bi{r}_{c}| $}, 
we calculated the matrix elements (\ref{eq:umat}) for $0 \le a \le 7$.
The number of points for the trapezoidal formula is $n=3\times 2^{11}$ and 
terms for the Fourier series (\ref{eq:fourier}) $M=161$.
Here $n$ is multiple of 3 
to make the set of points for the numerical integration closed in the translation in $t$ by $T/3$. 
The estimated error in numerical integration is less than $10^{-9}$ and 
lower twenty eigenvalues are obtained in 6 digits.

\subsection{Morse index and eigenfunctions for $a=1$}
\label{sec:morsehomo}
\begin{figure}
   \centering
   \includegraphics[width=16cm]{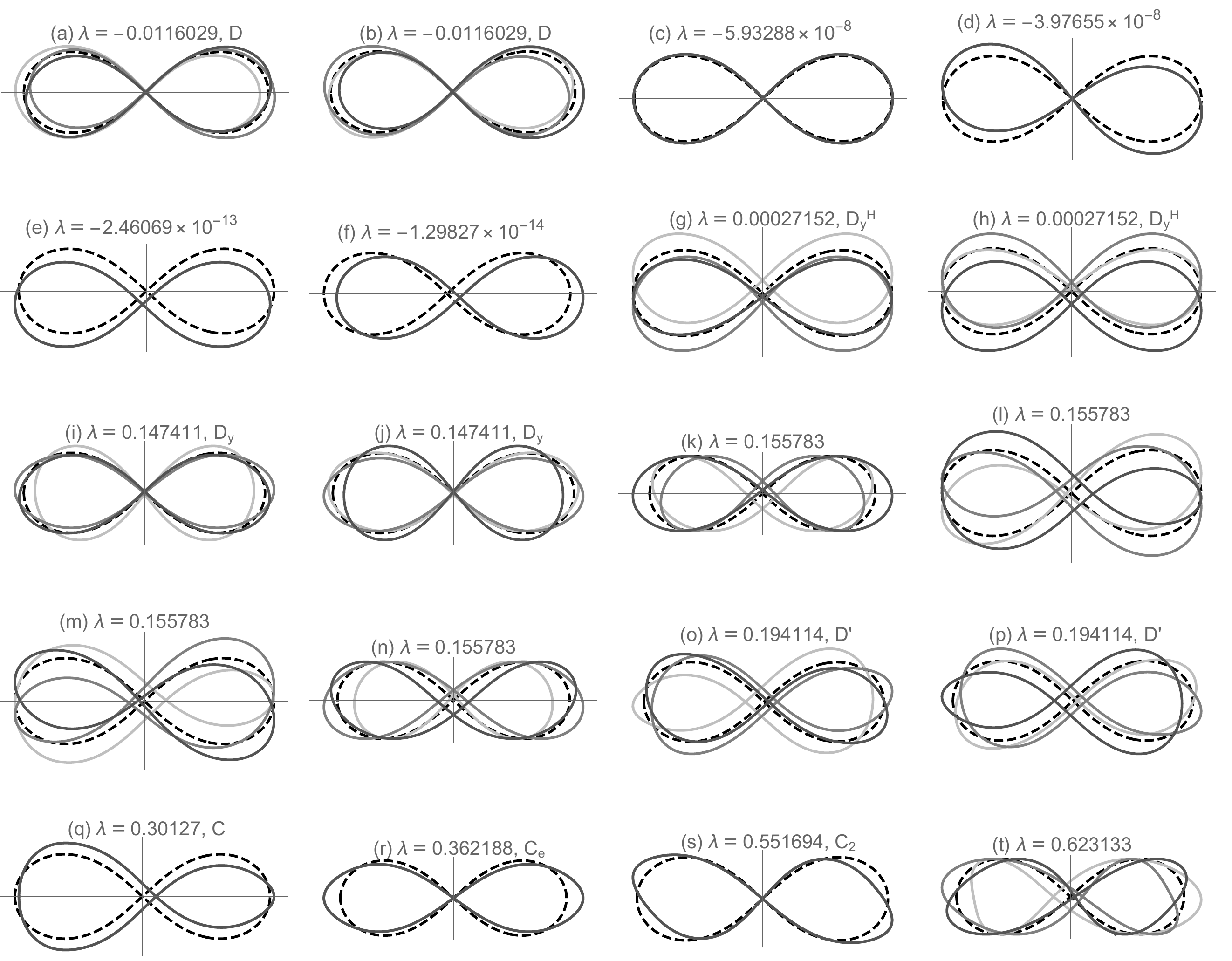} 
   \caption{
The eigenvalues, orbits 
\rev{$\bi{r}_1$} 
(dashed curve) for $a=1$, $x_{\max}=2$, $T=T_1=15.919135$, $S=33.225363$ 
and the variated orbits
\rev{$\bi{r}_b+h\delta\bi{r}_b$, $b=1,2,3$} 
(
\rev{light, medium and dark gray curves, respectively}), 
\rev{$h=1.5$}. 
\rev{Symbols at the end of the labels are those of eigenfunctions explained in table~\ref{tb:corr}.}
}
   \label{fig:8all}
\end{figure}

In figure~\ref{fig:8all}, for $a=1$,
twenty eigenvalues and eigenfunctions for 
the figure-eight choreography $q$ with size 
$x_{\max}=\max{x_{\rev{b}}(t)}=2$ 
and period $T=T_1=15.919135$
are shown in the ascending order from the minimum eigenvalue.
In figure~\ref{fig:8all}, the eigenfunction $\psi$ itself is not shown 
but the variated orbit 
\rev{$\bi{r}_{b}+h \delta \bi{r}_b$, $b=1,2,3$} with 
\rev{$h=1.5$} 
is displayed by 
\rev{light, medium and dark gray} curves, \rev{respectively,} together with the orbit 
$\rev{\bi{r}_1}$ 
by dashed curve,
\rev{where $\delta\bi{r}_b$ is the body $b$ component of $\psi$ defined by}
\begin{equation}\label{eq:dr}
	\rev{\delta\bi{r}_b=(\psi_{2b-1},\psi_{2b})^*}.
\end{equation}
The variated orbits are more physical and convenient 
for understanding the characteristics of eigenfunctions
though they include a parameter $h$ 
\rev{than eigenfunction itself}. 

The four eigenvalues in figure~\ref{fig:8all} (c) to (f) are close to zero and 
from the variated orbits we can see that they represent 
the translation in time, rotation, and translation in $x$- and $y$-direction,
respectively. 
Thus they are four zero eigenvalues originated in the conservation law.
The quadruply degenerated eigenvalues $\lambda=0.155783=\omega^2$ 
in figure~\ref{fig:8all} (k) to (n) are trivial.
Also the $\lambda=0.623133=4 \omega^2$ in figure~\ref{fig:8all} (t) is
one of the quadruply degenerated trivial eigenvalues with $k=2$.

The eigenvalues in figure~\ref{fig:8all} (q), (r) and (s) are non-degenerated. 
We show if an eigenvalue is non-degenerated \rev{like that} its eigenfunction is choreographic,
thus they are choreographic.
Suppose $\lambda$ is non-degenerated eigenvalue and $\psi$ its eigenfunction.
Since $\hat{H}$ and $\hat{C}$ commute, 
$\hat{C}\psi$ is also the eigenfunction of $\lambda$, 
thus $\hat{C}\psi=c \psi$ where $c$ is a real coefficient.
Then $\hat{C}^3=1$ leads $c^3=1$ and $\hat{C}\psi=\psi$ 
which means $\psi$ is choreographic. 

For choreographic eigenfunction $\psi$ the three variated orbits
overlap and differ only in time shift,
\begin{equation*}\label{key}
	\rev{\bi{r}_{c}(t)+h \delta\bi{r}_{c}(t)=\bi{r}_b(t')+h\delta\bi{r}_b(t'), \; t'=t+(c-b)T/3},
\end{equation*}
by $\hat{C}q=q$ and $\hat{C}\psi=\psi$ with (\ref{eq:Chat}).
Thus full curves in figure~\ref{fig:8all} (q), (r) and (s) overlap and appear as one.
Further, among the three choreographic eigenfunctions 
only the variated orbit (r) is symmetric in both $x$- and $y$-axis, 
thus it is the only figure-eight choreographic eigenfunction.  

The pair of successive eigenvalues 
(a) and (b), 
(g) and (h), 
(i) and (j), and 
(o) and (p) in figure~\ref{fig:8all} 
are doubly degenerated
and their variated orbits are splited into distinct full curves.
We show that any linear combination of such degenerated eigenfunctions
can not be choreographic, thus they are non-choreographic.
%
Suppose $\psi^{(\rev{s})}$ and $\psi^{(\rev{s}+1)}$ are the exactly doubly degenerated 
orthonormal eigenfunctions having distinct full curves.
Thus 
$\hat{C}\psi^{(\rev{s})} \ne \psi^{(\rev{s})}$ and $\hat{C}\psi^{(\rev{s}+1)} \ne \psi^{(\rev{s}+1)}$ 
since $\hat{C}q = q$.
Since 
the operator $\hat{C}$ commutes with $\hat{H}$, 
conserves inner product as $(\hat{C}f, \hat{C}g)=(f,g)$,
and $\hat{C}^3=1$,  
the $\hat{C}$ is represented as $\theta_c=\pm 2\pi/3$ rotation
\begin{equation}\label{eq:C}
	R(\theta_c)=\left( \begin{array}{r r}
		\cos \theta_c & -\sin \theta_c \\
		\sin \theta_c & \cos \theta_c
	\end{array}\right)
\end{equation}
in the base functions $\psi^{(\rev{s})}$ and $\psi^{(\rev{s}+1)}$.
Here the sign of $\theta_c$ is fixed by the phase of the base functions.
Thus for any linear combination $\psi=h_1\psi^{(\rev{s})}+h_2\psi^{(\rev{s}+1)}$,
$\hat{C}\psi=\psi$ represented by $R(\theta_c) (h_1, h_2)^* = (h_1, h_2)^*$ 
leads $(h_1,h_2)=0$ which means $\psi$ can not be choreographic.


Now we can count three kind of Morse index for $a=1$ in the different domains  
from figure~\ref{fig:8all}. 
Since there are two negative eigenvalues, (a) and (b)  
in figure~\ref{fig:8all}, 
Morse index $N$ is counted as 2. 
They are doubly degenerated and 
have distinct full curves,
therefore $N_c$ for choreographic 
and $N_e$ for figure-eight choreographic domain 
are both counted as 0.

\subsection{Morse index for $a \ge 0$}

\begin{figure}
	\centering
	\includegraphics[width=6cm]{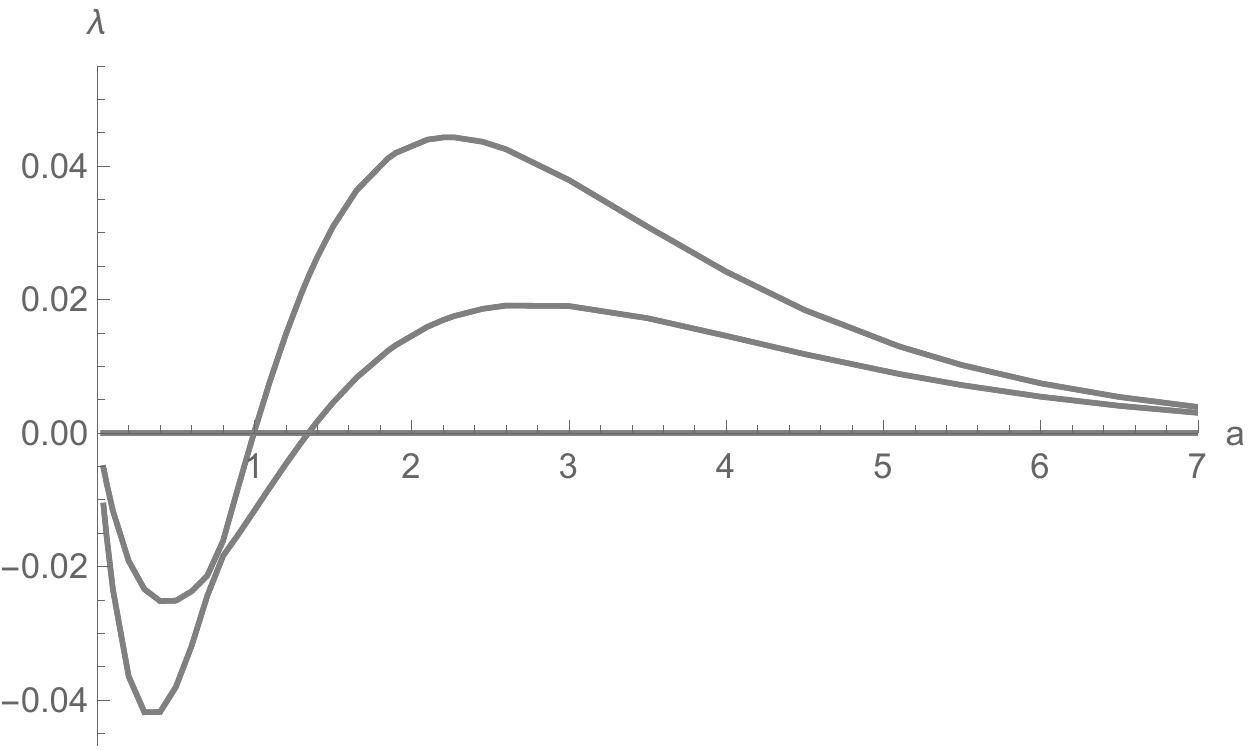} 
	\caption{
		The lowest eight eigenvalues $\lambda$'s for eigenvalue problem (\ref{eq:eigenBmat})  for homogeneous 
		potential with $0<a\le7$. }
	\label{fig:lambda-al}
\end{figure}
In figure~\ref{fig:lambda-al}, 
for $0<a\le7$, 
the lowest eight eigenvalues 
for the figure-eight choreography with the same size $x_{\max}=2$
are plotted as functions of $a$.
Two curves are doubly degenerated and there are four lines on the $x$-axis 
which are four zero eigenvalues.

The Morse indices for $0 \le a$ are
\begin{equation}\label{eq:N(a)}
\rev{N} = 
\left\lbrace \begin{array}{l l}
4 & (0 \le a < a_0), \\
2 & (a_0 < a < a_1), \\
0 & (a_1 < a),
\end{array}\right. 
\end{equation}
and
\begin{equation}\label{key}
  \rev{N_c}=\rev{N_e}=0 \; (0 \le a),
\end{equation}
where
\begin{equation}\label{key}
  a_0=0.9966, \; a_1=1.3424.
\end{equation}
Here $a=0$ is calculated by the log potential 
\begin{equation}\label{key}
  U(q)= \sum_{\rev{b>c}}\log r_{\rev{b c}}
\end{equation}
and $a>7$ are extrapolated.

The characteristics of 
the variated orbits for $a \ne 1$ 
are almost similar for $a=1$ shown in figure~\ref{fig:8all} 
though the order of the eigenvalues may be changed.
For example, at $a=1.5$, 
the first four eigenvalues are zero and
the fifth and the sixth variated orbits are similar to the first and the second in figure~\ref{fig:8all},
as read in figure~\ref{fig:lambda-al}.
We present precise 
table for \rev{characteristics of} the variated orbits 
in section \ref{sec:corr}.

\subsection{Sim\'{o}'s H orbits}
\label{sec:H}
The three orbits, H1, H2 and H3, 
found by Sim\'{o} \cite{simoH} are
very close to the variated orbits by non-choreographic eigenfunctions
for $a=1$. 
\begin{figure}
	\centering
	\includegraphics[width=6cm]{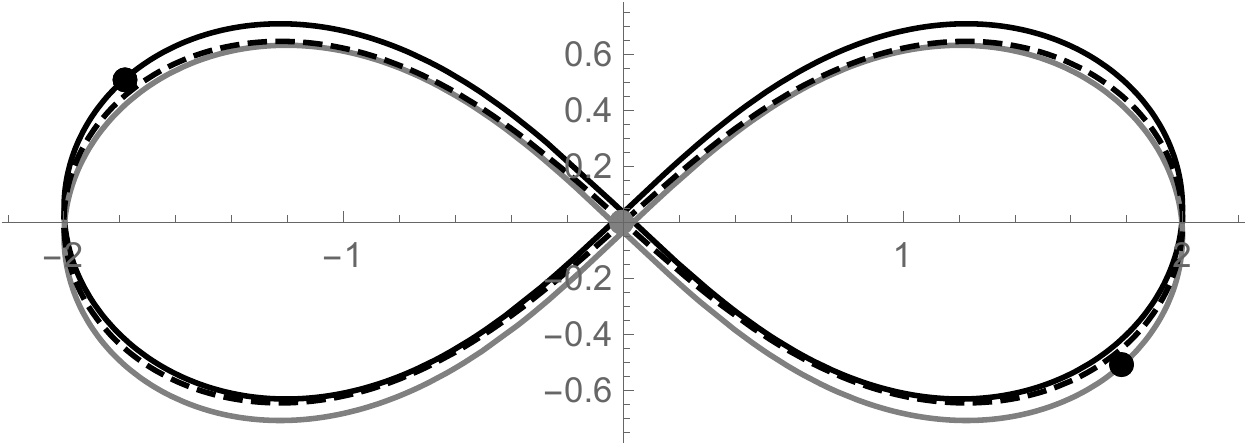} 
	\includegraphics[width=2.2cm]{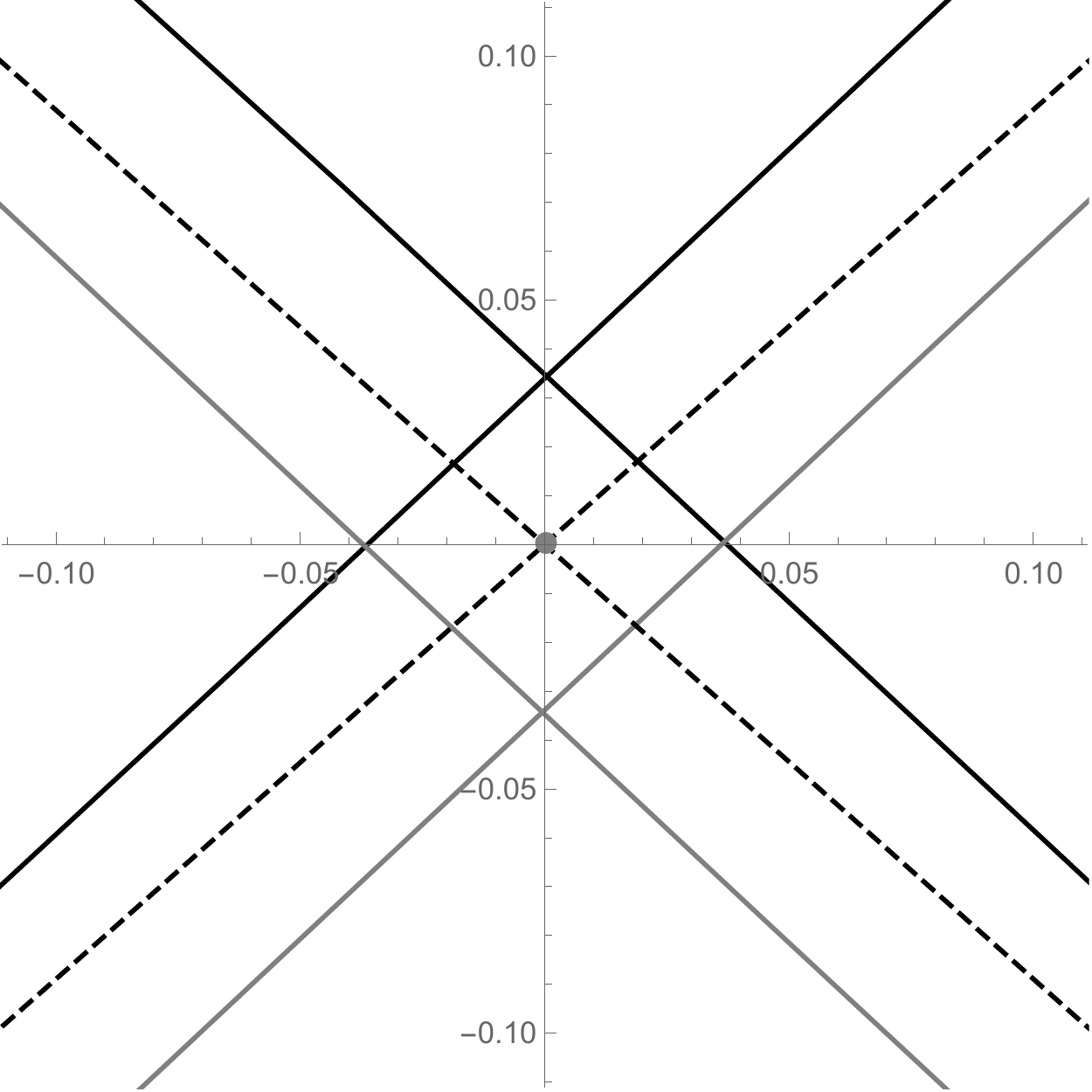}
	\\
	\hspace*{1.5cm} (a) \hspace*{3.5cm} (b) 
	\caption{
		(a) Sim\'{o}'s H3 orbit \cite{simoH} rotated by 0.277217~rad 
		scaled to unit mass $m=1$ and $T=T_1=15.919135$.
		(b) The same figure around the origin.}
	\label{fig:simoH3}
\end{figure}
In figure~\ref{fig:simoH3}, 
the H3 orbit rotated and scaled to $T=T_1$, $q^{H3}(t)$, are shown. 
It consists of three slightly different eight shaped orbits, 
one of them is passing through the origin as shown in figure~\ref{fig:simoH3} (b).
The set of orbits is symmetric in both $x$ and $y$ inversion
where the two orbits are exchanged in $y$ inversion.
The orbits H1 and H2 are the same orbit as H3 by rotation, translation in time and permutation of bodies.

The variated orbit 
\begin{equation}\label{eq:qTheta}
  q(t) + h \psi^{(\Theta)}(t)
\end{equation}
by doubly degenerated $\rev{s}$'th and $(\rev{s}+1)$'th eigenfunctions
$\psi^{(\rev{s})}$ and $\psi^{(\rev{s}+1)}$, 
\begin{equation}\label{eq:Theta}
  \psi^{(\Theta)}(t)  = \cos\Theta\psi^{(\rev{s})}(t)+\sin\Theta\psi^{(\rev{s}+1)}(t), 
\end{equation}
is very close to $q^{H3}$ at $\rev{s}=7$, $h=h_H=0.28375$ and some $\Theta=\Theta_H$.
Actually the squared difference between $q^{H3}$ and $q + h_H \psi^{(\Theta_H)}$
averaged in $t$
is less than $10^{-7}$.
The variated orbits for $\psi^{(7)}$ and $\psi^{(8)}$ 
are shown in figure~\ref{fig:8all} (g) and (h).
%
They are symmetric in $y$-axis but not in $x$-axis.
The linear combination (\ref{eq:Theta}) with $\Theta=\Theta_H$ makes it symmetric in $x$-axis.

\begin{figure}
	\centering
	\includegraphics[width=6cm]{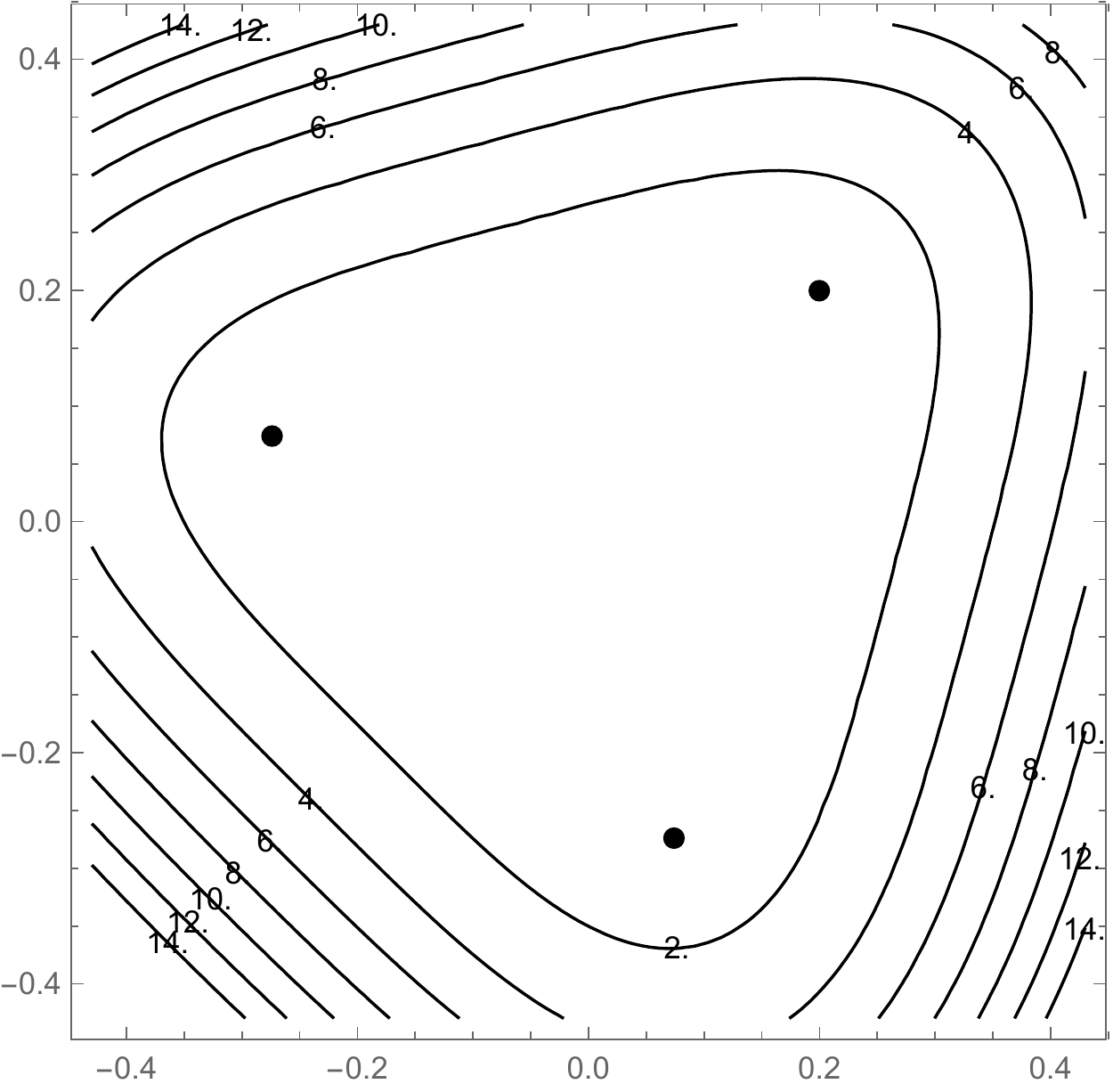} 
	\caption{
    Contour plot for action $S(q+h \psi^{(\Theta)})$ where horizontal and vertical axis are $h\cos\Theta$ and $h\sin\Theta$, respectively. 
    Contours are labeled by $(S(q+h \psi^{(\Theta)})-S(q)) \times 10^5$.
    }
	\label{fig:simoH3Scont}
\end{figure}
In figure~\ref{fig:simoH3Scont}, the contour plot of action $S(q+h \psi^{(\Theta)})$ is shown
where horizontal and vertical axis are $h\cos\Theta$ and $h\sin\Theta$, respectively. 
The contours in figure~\ref{fig:simoH3Scont} show the three fold symmetry  
since $\hat{C}$ is represented by $\theta_c=\pm 2\pi/3$ rotation (\ref{eq:C}) and  
conserves $S$ as $S(\hat{C}f)=S(f)$.

One of the three black points in figure~\ref{fig:simoH3Scont} is 
the point $(h_H, \Theta_H)$
closest to the critical point $q^{H3}$ of action functional, 
and the other two $(h_H, \Theta_H \pm 2\pi/3)$ 
its cyclic permutations of bodies with time shift, 
$\hat{C} q^{H3}$ and $\hat{C}^2 q^{H3}$.

The action \rev{$S(q^{H3})=33.22536589$} at $q^{H3}$ 
is slightly higher than \rev{$S(q)=33.22536229$} at $q$, 
thus the critical point $q^{H3}$ will be 
\rev{local maximum} 
towards $q$.
\rev{Here $S(q^{H3})$ and $S(q)$ are obtained in multiple precision calculation by the initial conditions in \cite{simoH}, and}
no shallow 
\rev{local maximum} 
is found numerically in the plane around the black points
in figure~\ref{fig:simoH3Scont}.
Nevertheless
$h_H$ is about $0.282$ estimated by
\begin{equation}\label{eq:posbylambda}
  h=\sqrt{\frac{6 |S(q')-S(q)|}{|\lambda|}}
\end{equation}
with $q'=q^{H3}$, which assumes critical at $h$.  
The estimation (\ref{eq:posbylambda}) is derived by equating $S(q')$ and 
the $S(q+h\psi)$ truncated at $h^3$ term in (\ref{eq:S}) regarding $S^{(3)}$ as a parameter,
$
S(q)+h^2 \lambda/2 + h^3 S^{(3)}/3!,
$
with the critical condition at $h$, $h \lambda+h^2 S^{(3)}/2=0$.

Note that Sim\'{o}'s $q^{H3}$ approximately satisfies the relation \cite{simoH}
\begin{equation} \label{eq:H3andq}
  q_i(t) \simeq \frac{1}{3} \sum_{k=0}^2 q_{i+2 k}^{H3}(t-\frac{k T}{3})
\end{equation}
or
\begin{equation}\label{key}
  q \simeq \frac{1+\hat{C}+\hat{C}^2}{3} q^{H3} 
\end{equation}
since $q+h_H \psi^{(\Theta_H)}$ at $\rev{s}=7$ is very close to $q^{H3}$
and 
\begin{equation} \label{eq:sumC}
  (1+\hat{C}+\hat{C}^2) \psi^{(\Theta)}=0
\end{equation}
by equation (\ref{eq:C}).

\section{Lennard-Jones-type potential}
\label{sec:LJ}
\begin{figure}
	\centering
	\includegraphics[width=10cm]{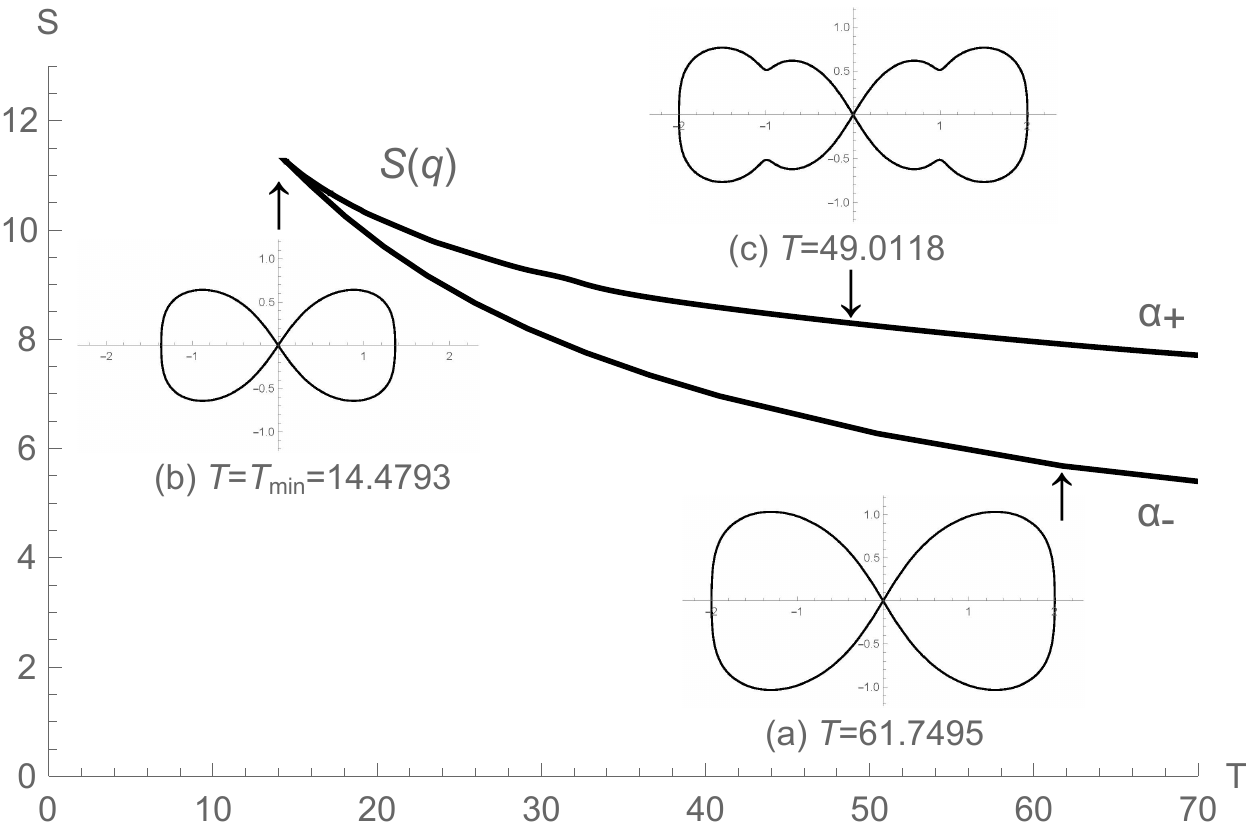} 
	\caption{
		Action $S(q)$ for the figure-eight solution $\alpha$ against $T$ and 
		the orbit $q$ under LJ potential. 
		(a) Figure-eight orbit with $x_{\max}=2$ ($T=61.7495$) in $\alpha_-$ branch.
		(b) Figure-eight orbit at $T=T_{\min}=14.4793$, branch point.
		(c) Gourd shaped orbit with $x_{\max}=2$ ($T=49.0118$)  in $\alpha_+$ branch.
		The three orbit, (a)--(c) are drawn in relatively correct size.
	}
	\label{fig:ap2ed}
\end{figure}

For the system interacting through the LJ potential
\begin{equation}\label{key}
  U(q)=\sum_{\rev{b>c}}u^{LJ}(r_{\rev{b c}}),
\end{equation}
we calculated the Morse index of the solution $\alpha$ \cite{fukuda}. 
The solution $\alpha$ is the figure-eight choreographic solution 
asymptotically tending to that under homogeneous potential with $a=6$ at $T \to \infty$.
In figure~\ref{fig:ap2ed}, $S(q)$ for the solution $\alpha$ is shown against $T$. 
There are two branches 
\rev{of $S(q)$} branched at $T=T_{\min}=14.4793$, 
the minimum period $T$ of solution $\alpha$, 
as shown in figure~\ref{fig:ap2ed}.
We denote the branch with higher action value $S$ as the $\alpha_+$ 
and lower the $\alpha_-$. 

The shape of the orbit gradually changes from the figure-eight 
for $\alpha_-$ shown in figure~\ref{fig:ap2ed} (a), 
via the branch point $T=T_{\min}$ in figure~\ref{fig:ap2ed} (b),
to the gourd shape for $\alpha_+$ shown in figure~\ref{fig:ap2ed} (c).
Though the $S(q)$ shows cusp like shape at $T=T_{\min}$, $q$ changes smoothly there.  
%
\rev{
The characteristics of the eight-shaped choreographic orbits 
of $\alpha_-$ for $T \to \infty$
are very close to those for the $a=6$ homogeneous potential 
since the particles have large relative distances and 
the short-range repulsive part of the LJ potential is less important \cite{fukuda}.}

\begin{figure}
	\centering
	\includegraphics[width=12cm]{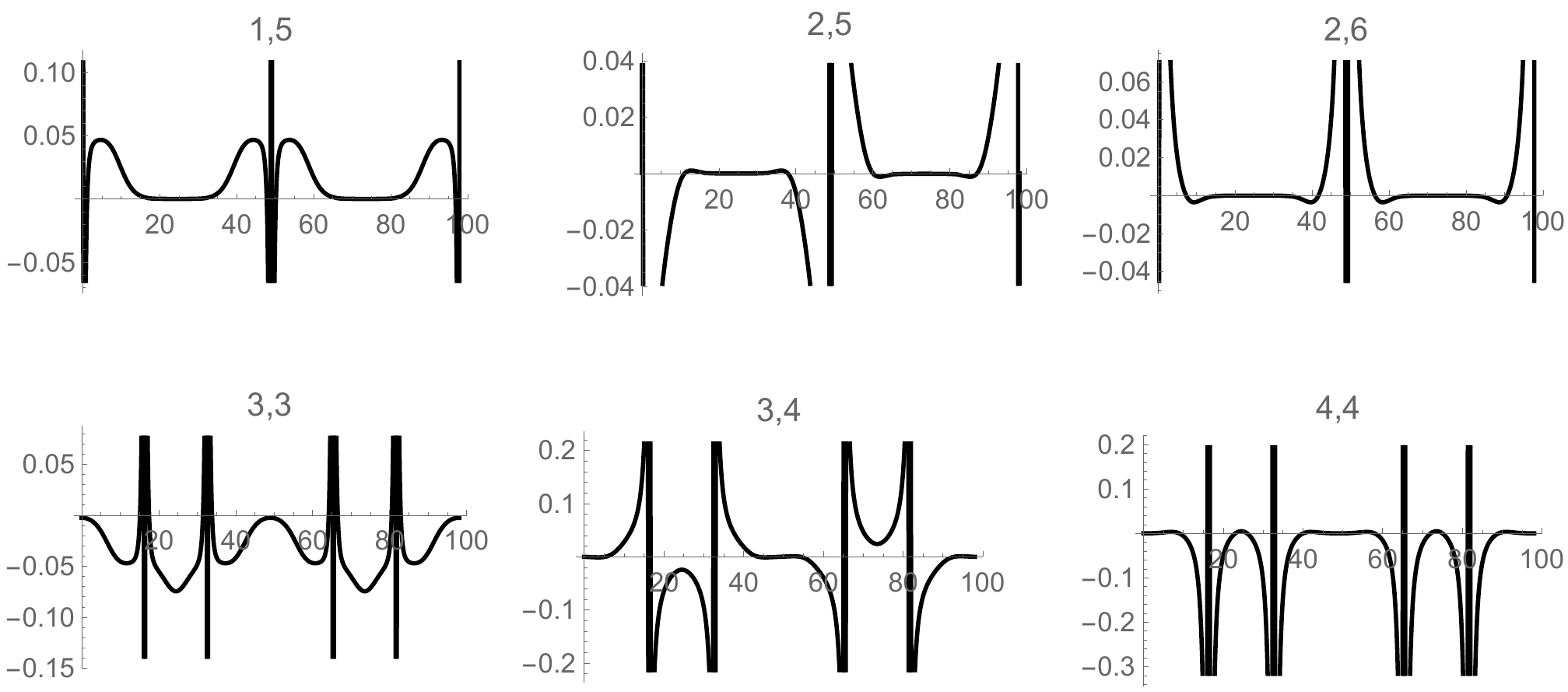} 
	\caption{
		Independent elements of $U_{i j}(t)$ for solution $\alpha_+$ at $T=98.0332$
		($x_{\max}=2.4$). 
		Each graph is titled by $i,j$.
	}
	\label{fig:ap120u}
\end{figure}

The numerical calculations are done 
with $3\times2^{11} \le n \le 3\times 2^{14}$ and 
$321 \le M \le 10241$
for $T_{\min} \le T < 100$, 
and lower twenty eigenvalues are obtained at least 5 digits.
For the solution $\alpha_+$ the gourd shape is sharper for larger $T$ 
and sharp spikes appear in the matrix elements $U_{i j}(t)$
as shown in figure~\ref{fig:ap120u}, 
which make the numerical integration for the $\alpha_+$ for $T > 100$ difficult.

We obtain for $\alpha_-$
\begin{equation}
\rev{N}(\alpha_-) = \left\lbrace \begin{array}{l l}
5 & (14.4793 \le T < 14.5952), \\
4 & (14.5952 < T < 14.8358), \\
\rev{2} & \rev{(14.8358 < T < 14.8611),} \\
0 & (\rev{14.8611} < T), \\
\end{array} \right. 
\end{equation}
\begin{equation}
N_c(\alpha_-) = \left\lbrace \begin{array}{l l}
1 & (14.4793 \le T < 14.5952), \\
0 & (14.5952 < T), \\
\end{array} \right.
\end{equation}
\begin{equation}\label{eq:Nep}
  N_e(\alpha_-) = 0 \; (14.4793 \le T),
\end{equation}
and for $\alpha_+$
\begin{equation}\label{eq:N}
\rev{N}(\alpha_+) = \left\lbrace\begin{array}{l l}
6 & (14.4793 \le T < 16.1110), \\
8 & (16.1110 < T < 16.8779), \\
10 & (16.8687 < T < 17.1317), \\
11 & (17.1317 < T < 18.6154), \\
12 & (18.6154 < T),
\end{array}\right.
\end{equation}
\begin{equation}\label{eq:Nc}
\rev{N_c}(\alpha_+) = \left\lbrace\begin{array}{l l}
2 &  (14.4793 \le T < 17.1317), \\
3 & (17.1317 < T < 18.6154), \\
4 & (18.6154 < T),
\end{array}\right.
\end{equation}
\begin{equation}\label{eq:Nem}
  \rev{N_e}(\alpha_+) =1 \;  (14.4793 < T).
\end{equation}
For $T > 100$ all indices are extrapolated.

\begin{figure}
	\centering
	\includegraphics[width=8cm]{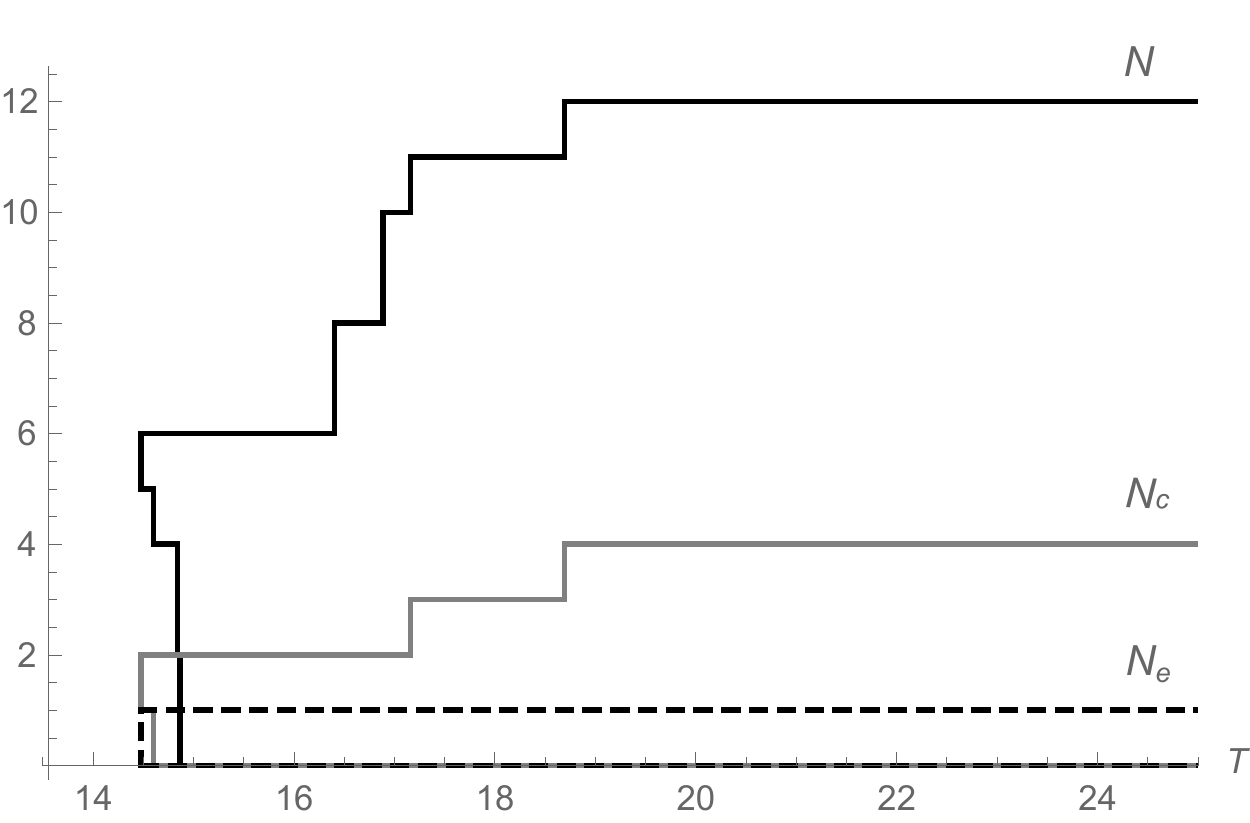} 
	\caption{
		$N$, $N_c$ and $N_e$ for the solution $\alpha$.
	}
	\label{fig:alphaIndexTed}
\end{figure}

In figure~\ref{fig:alphaIndexTed}, \rev{the} indices $N$, $N_c$ and $N_e$ are plotted 
against  $T$ for $\alpha_+$ and $\alpha_-$ together.
All \rev{the} $N$, $N_c$ and $N_e$, increase from $0$ monotonically 
to $12$, $4$ and $1$, respectively, starting from infinitely large $T$ in $\alpha_-$.
They all jump by one at $T=T_{\min}$.
    
\subsection{Correlation of eigenfunctions}
\label{sec:corr}
\begin{figure}
	\centering
	\includegraphics[width=16cm]{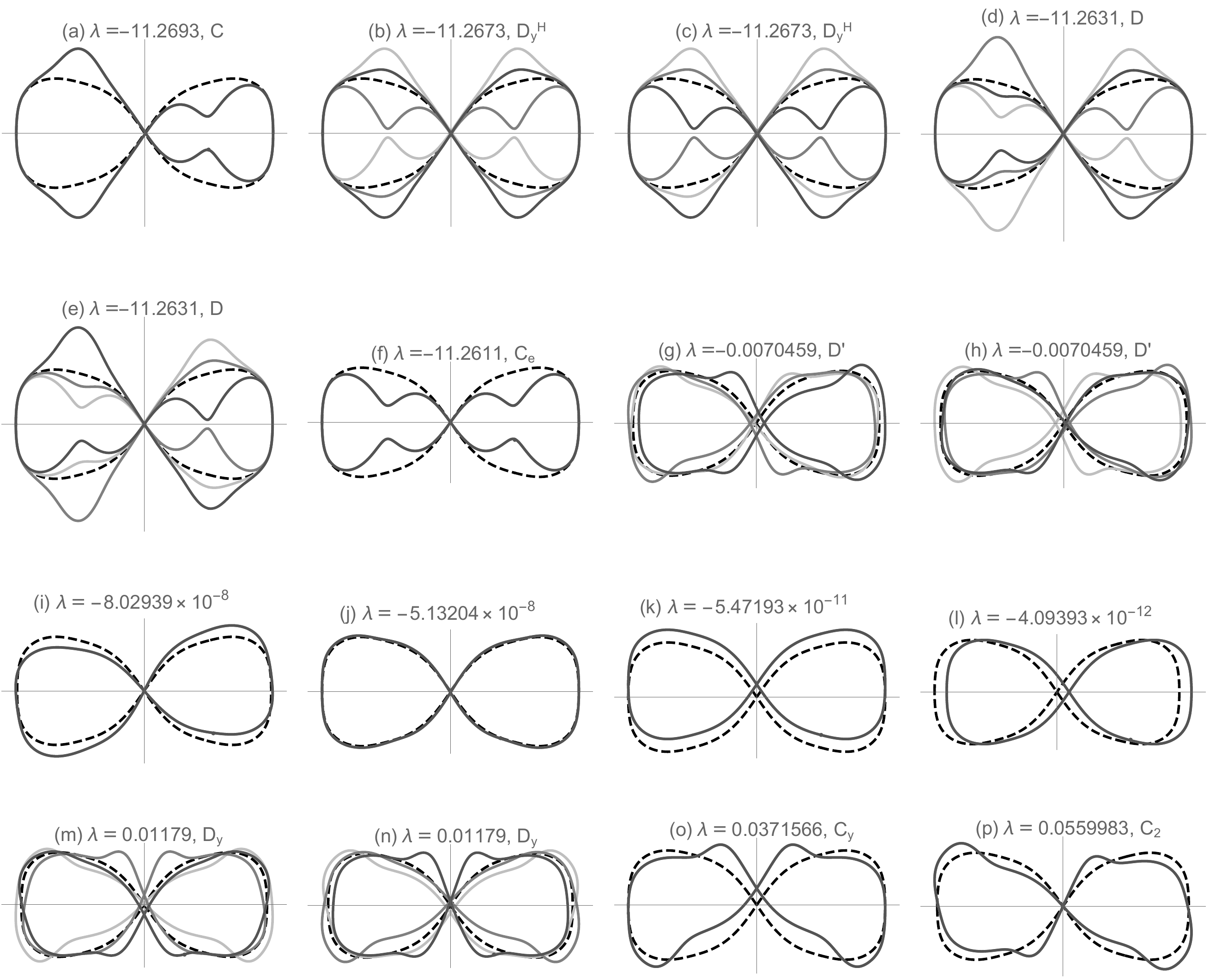} 
	\caption{
		The eigenvalues, orbits 
		\rev{$\bi{r}_1$}
		(dashed curve) for solution $\alpha_+$ under LJ potential, 
		$x_{\max}=1.44$, $T=16.4019$, $S=10.8136$, 
		and the variated orbits 
		\rev{$\bi{r}_b+h\delta\bi{r}_b$, $b=1,2,3$}
		 (\rev{light, medium and dark gray} 
		 curves, respectively) with $h=1$.
\rev{Symbols at the end of the labels are those of eigenfunctions explained in table~\ref{tb:corr}.}
	}
	\label{fig:ap072pub}
\end{figure}
In figure~\ref{fig:ap072pub}, 
lower sixteen eigenvalues and eigenfunctions for the solution $\alpha_+$ at 
$T=16.4019$ 
are shown 
\rev{in the same style} 
as
figure \ref{fig:8all}.
Instead of exhibiting a huge number of similar lists as figure~\ref{fig:ap072pub} 
for different $T$,
we make figure~\ref{fig:ap072pub} representative and show 
\rev{how they change.
When the period $T$ changes, either in the $\alpha_+$ or $\alpha_-$ branch, 
the different eigenfunctions change continuously with $T$.}
%
%
\begin{figure}
	\centering
	\includegraphics[width=3.5cm]{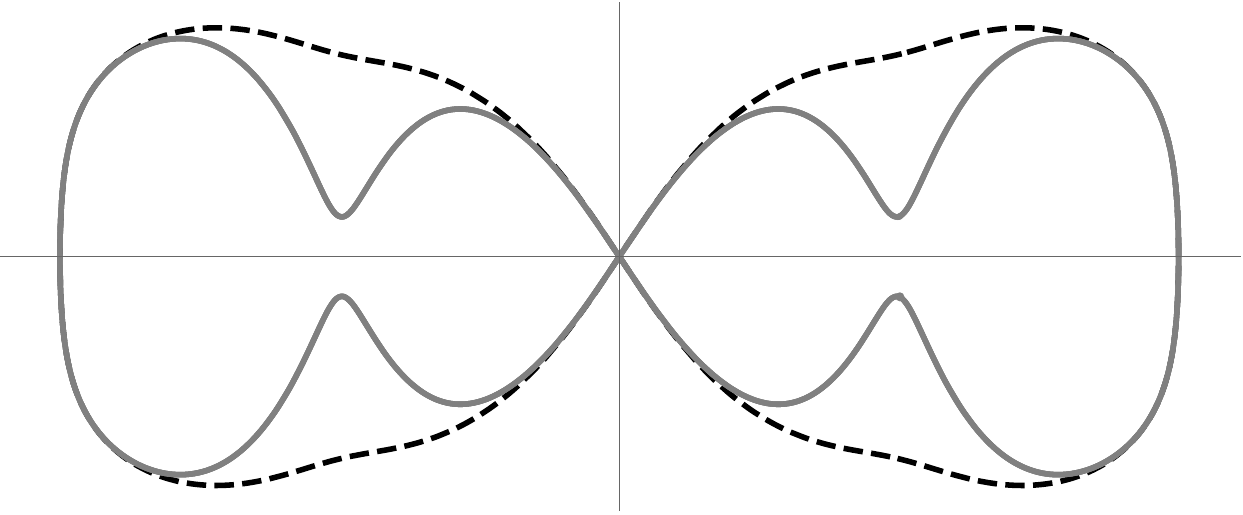} 
	\includegraphics[width=3.5cm]{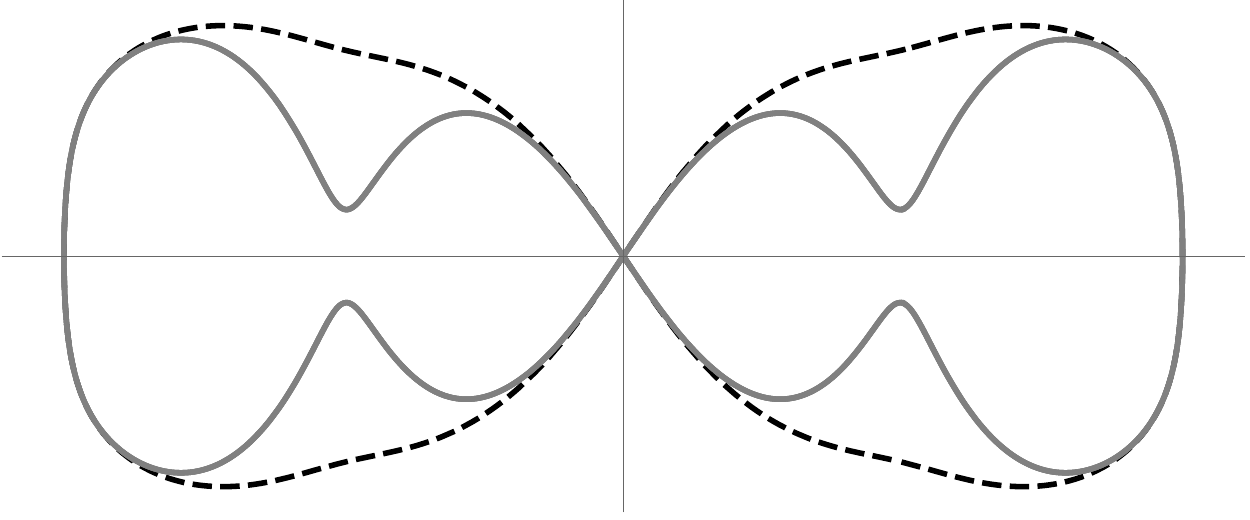} 
	\includegraphics[width=3.5cm]{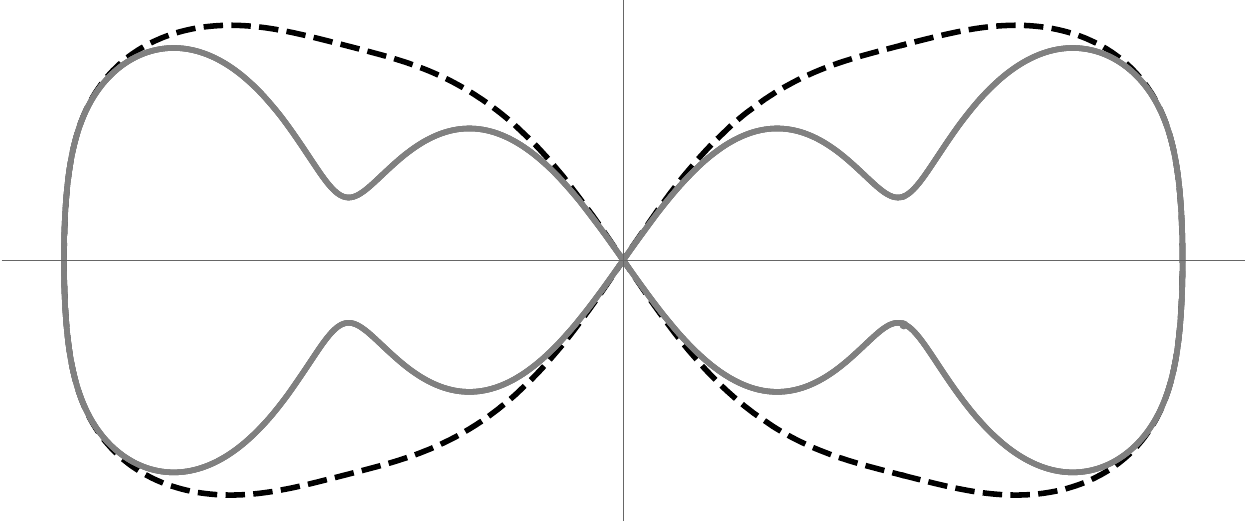} 
	\includegraphics[width=3.5cm]{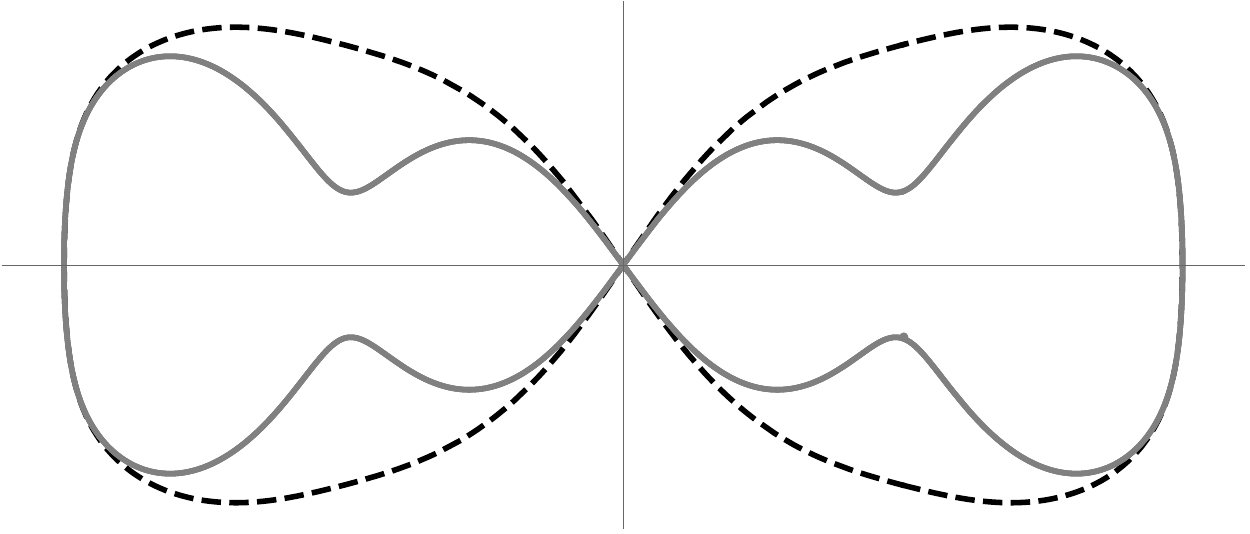} 
	\\
	(a) $T=19.0588$ \hspace{0.5cm} 
	(b) $T=18.3370$ \hspace{0.5cm} 
	(c) $T=17.0085$ \hspace{0.5cm} 
	(d) $T=16.4019$
	\\
	\includegraphics[width=3.5cm]{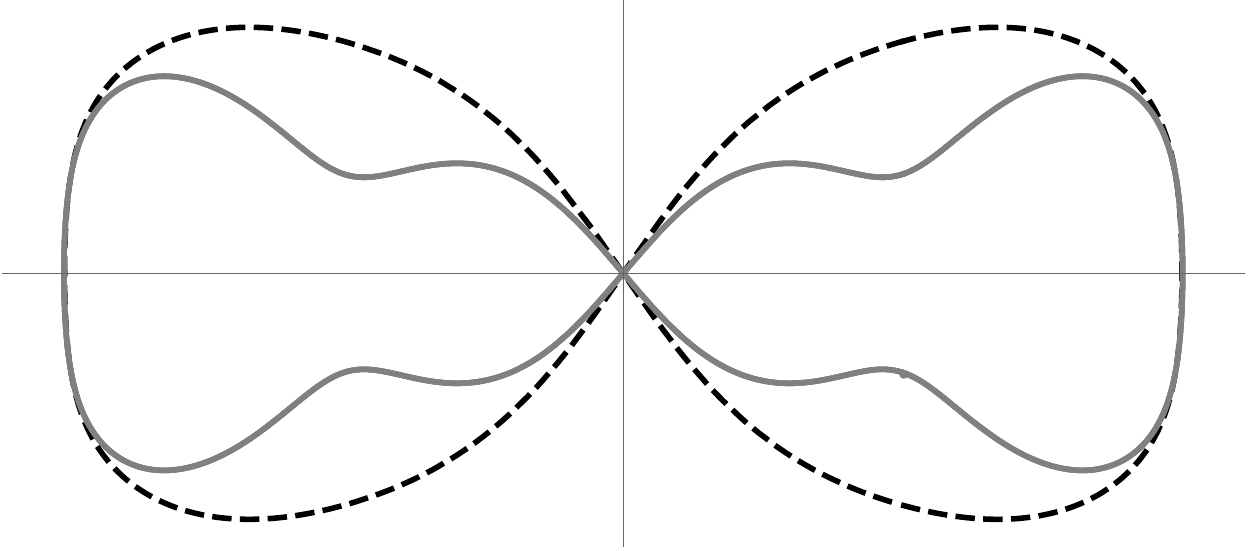} 
	\includegraphics[width=3.5cm]{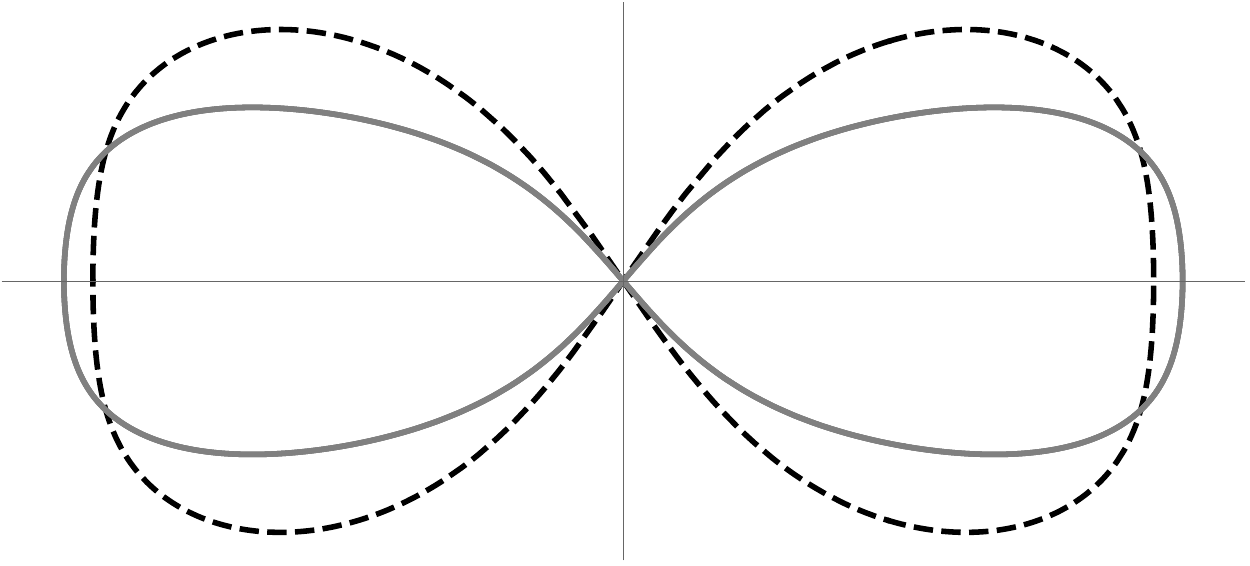} 
	\includegraphics[width=3.5cm]{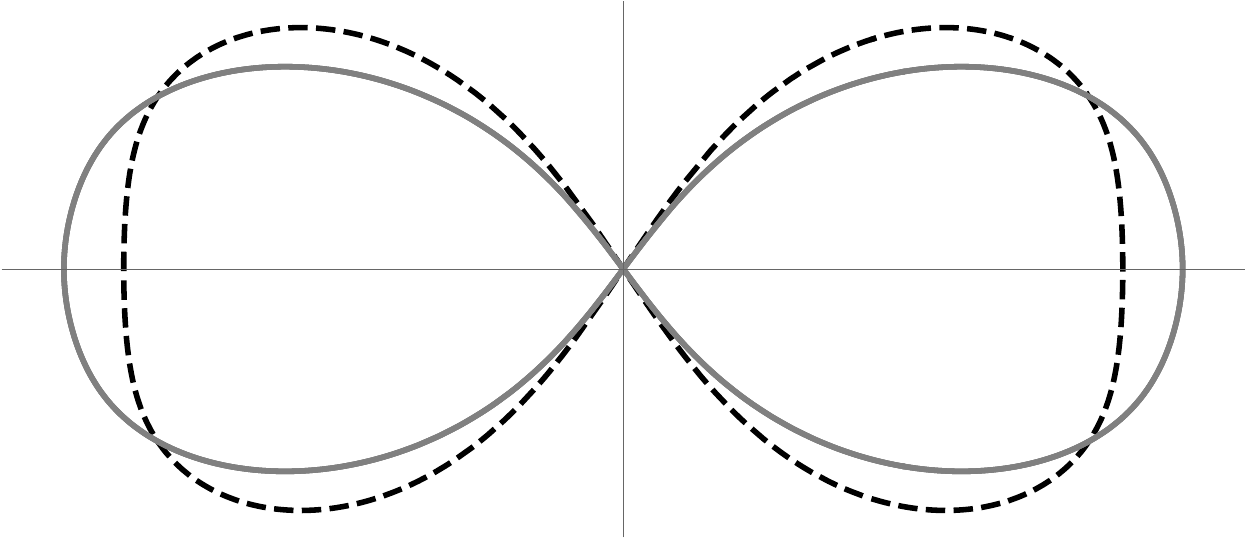} 
	\includegraphics[width=3.5cm]{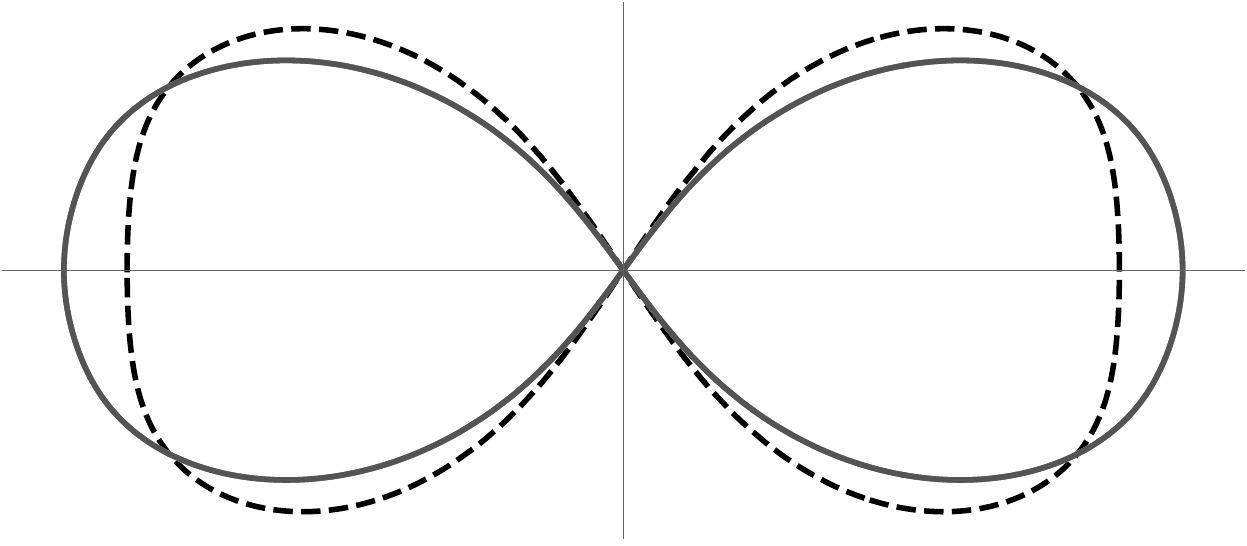} 
	\\
	(e) $T=15.3047$ \hspace{0.5cm} 
	(f) $T=14.4869$ \hspace{0.5cm} 
	(g) $T=14.6763$ \hspace{0.5cm} 
	(h) $T=14.8420$ \hspace{0.5cm} 
	\\
	\includegraphics[width=3.5cm]{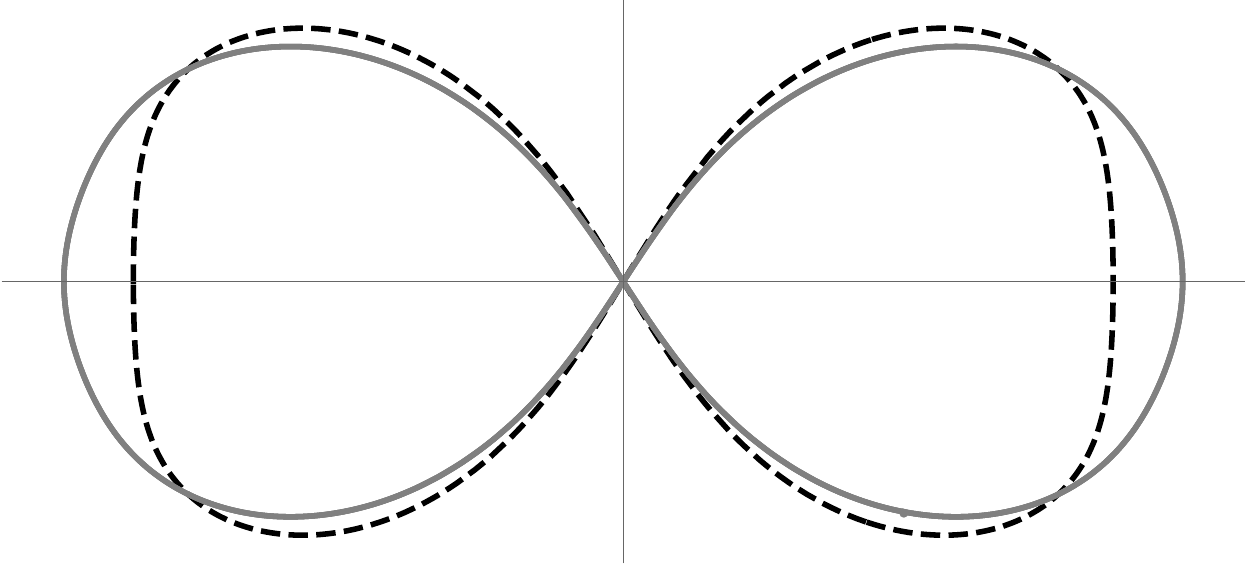} 
	\includegraphics[width=3.5cm]{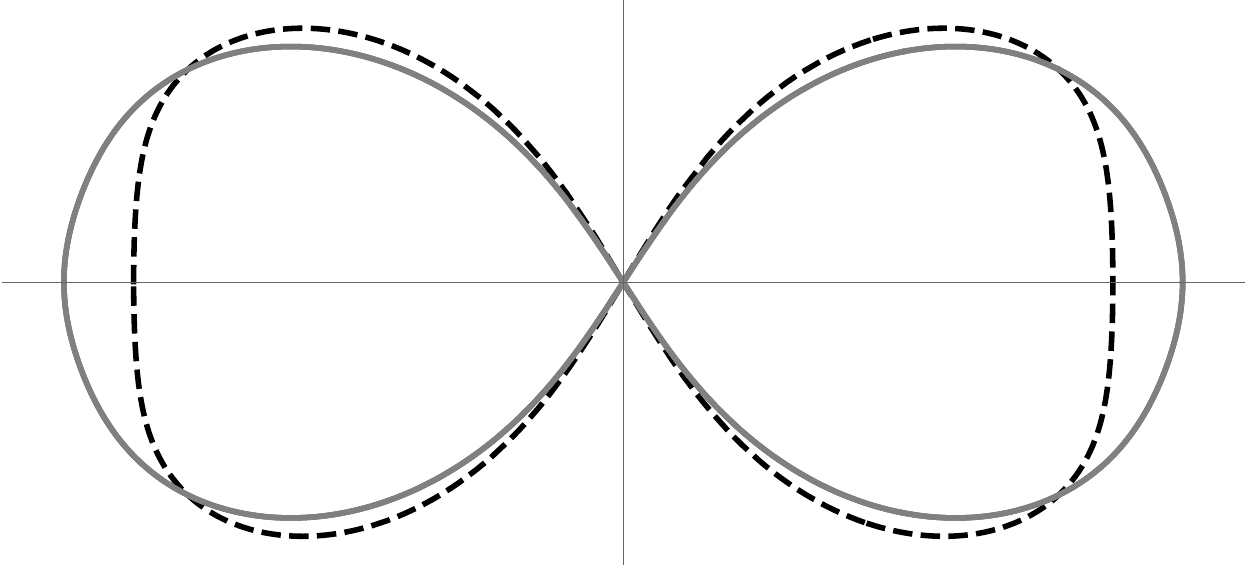} 
	\includegraphics[width=3.5cm]{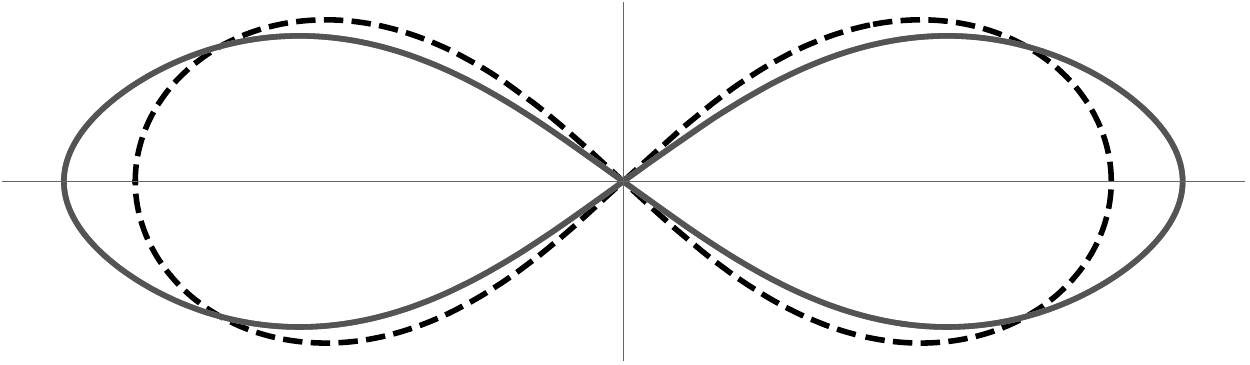} 
	\hspace*{3.5cm}
	\\
	(i) $T=61.7495$ \hspace*{1cm}
	(j) $a=6$ \hspace*{2cm} 
	(k) $a=1$ \hspace*{4cm} 
	\caption{
		The eigenvalues, orbits $\bi{r}_1$ (dashed curve) and the variated orbits $\bi{r}_b+h \delta \bi{r}_b$,
		$b=1,2,3$ (light, medium and dark gray curves, respectively) 
		correlated to those shown in figure \ref{fig:ap072pub} (f) or to  
		eigenfunction $C_e$ defined in table~\ref{tb:corr}, with
		$h=3$ for (i) and (j), $h=1.5$ for (k), and $h=1$ for the others.
	}
	\label{fig:cor}
\end{figure}
\rev{
For example, 
changes of figure~\ref{fig:ap072pub} (f) are shown
in figure \ref{fig:cor}.  
Starting from the figure~\ref{fig:cor} (d) which is figure~\ref{fig:ap072pub} (f),
figures~\ref{fig:cor} (a)--(e) show continuous changes in the $\alpha_+$ branch
and figures~\ref{fig:cor} (f)--(i) 
in the $\alpha_-$ branch.}
\par
\rev{
Since the $\alpha_-$ solution for $T \to \infty$ tends to the solution for the $a=6$ homogeneous system,
their eigenfunctions and the variated orbits also do so.
In the case of figure~\ref{fig:cor}, 
(i) and (j) are very close 
since they are the variated orbits for the $\alpha_-$ for large $T$ 
and the $a=6$ homogeneous system, respectively. }

\rev{We call the eigenfunction obtained by changing continuous parameter, $T$ or $a$, \textit{correlated}. 
	We also call two eigenfunctions, one for $\alpha$ solution with $T$ and 
	the other homogeneous system with $a$,  
	if they are identical at $T \to \infty$ in the $\alpha_-$ branch and at $a=6$,
	\textit{correlated}.
In figure \ref{fig:cor}, the eigenfunctions of all variated orbits, (a)--(k), 
are correlated since (i) and (j) are correlated.
}

In table~\ref{tb:corr} 
\rev{eigenfunctions correlated to those} 
shown in figure~\ref{fig:ap072pub}, 
except for the trivial eigenfunctions, are tabulated.
Each row shows 
the eigenfunctions by symbols 
in ascending order of eigenvalues
for the solution shown in the left three columns.
The symbol 
$C$ represents non-degenerated choreographic eigenfunction, 
$D$ doubly degenerated non-choreographic eigenfunctions and
$0$ quadruply degenerated eigenfunctions of the zero eigenvalue. 

The subscript 
\rev{in the symbol} indicates \rev{the symmetry of}
the variated orbits\rev{. $y$ indicates they} 
are symmetric in $y$ axis,
$e$ in both $x$ and $y$ axis, and  
\rev{2 in 2 fold rotation at origin}. 
Prime 
\rev{are used to distinguish different} eigenfunctions
\rev{with the same characters}.
The superscript $H$ 
identifies the eigenfunction corresponding to the Sim\'{o}'s H obits 
discussed in section \ref{sec:H} for $a=1$,
(g) and (h) in figure~\ref{fig:8all}.

\rev{
These symbols defined here are added at the end of each label in figure~\ref{fig:8all} and \ref{fig:ap072pub}
except for 0's.
}
\rev{The variated orbits shown in figure~\ref{fig:cor} are those with the eigenfunction $C_e$.}

\begin{table}
	\centering
	\caption{
	The correlation table for eigenfunctions. 
	Each row shows the type of eigenfunctions 
	in ascending order of the eigenvalue. 
	$C$; a choreographic eigenfunction. 
	$D$; doubly degenerated non-choreographic eigenfunctions. 
	$0$; quadruply degenerated \rev{eigenfunctions of} zero eigenvalue. 
    Subscript $y$; the variated orbits are symmetric in $y$ axis.
	Subscript $e$; symmetric in both $x$ and $y$ axes. 
	\rev{Subscript 2; symmetric in 2 fold rotation at origin}. 
	Superscript $H$; the eigenfunction corresponds to Sim\'{o}'s H solution for $a=1$.
    Prime 
    eigenfunctions
\rev{with the same characters}.
    }
    \label{tb:corr}
	\begin{tabular}{c r r c c c c c c c c c}
	\hline
	 &  \multicolumn{1}{c}{$T$} & \multicolumn{1}{c}{$N$}  & \multicolumn{9}{l}{type of eigenfunctions} \\
	\hline
	$\alpha_+$  
	& 19.0588 &12 & $C$ & $D_y^H$ & $D$ & $C_e$ & $C_y$ & $D'$ & $D_y$ & $C_{\rev{2}}$ &  $0$ \\
	 & 18.3370 & 11 & $C$ & $D_y^H$ & $D$ & $C_e$ & $D'$ & $C_y$ & $D_y$ & $0$ & $C_{\rev{2}}$ \\
	 & 17.0085 & 10 & $C$ & $D_y^H$ & $D$ & $C_e$ & $D'$ & $D_y$    &$0$&  $C_y$   & $C_{\rev{2}}$ \\
	 & 16.4019 & 8   & $C$ & $D_y^H$ & $D$ & $C_e$ & $D'$  &$0$ & $D_y$ & $C_y$ & $C_{\rev{2}}$ \\
 	 & 15.3047 & 6   & $C$ & $D_y^H$ & $D$ & $C_e$ &$0$& $D'$    & $D_y$ & $C_{\rev{2}}$ & $C_y$ \\
	 \rev{$\alpha_-$} 
	 & 14.4869 & 5   & $D_y^H$ & $C$ & $D$ &$0$& $D_y$  & $D'$ & $C_e$ &  $C_{\rev{2}}$ & $D_y'$ \\ 
	 & 14.6763 & 4 & $D_y^H$ & $D$ &$0$& $C$ & $D_y$  & $D'$ & $C_e$ &  $C_{\rev{2}}$ & $D_y'$ \\
	& \rev{14.8420}  & \rev{2} & \rev{$D$} & \rev{$0$} & \rev{$D_y^H$} & \rev{$D_y$} & \rev{$C$} & \rev{$D'$}  & \rev{$C_e$} & \rev{$C_2$} & \rev{$C_y$} \\
	& \rev{61.7495}  & 0 & $0$& $D$ & $D_y^H$ & $D_y$ & $D'$  & $C_e$ & $C$ &  $C_{\rev{2}}$ & $C_y$ \\
	\multicolumn{2}{c}{$a=6$}  & 0 & $0$& $D$ & $D_y^H$ & $D_y$ & $D'$  & $C_e$ & $C$ &  $C_{\rev{2}}$ & $C_y$ \\
	\multicolumn{2}{c}{$a=1$}  & 2 & $D$ &$0$& $D_y^H$ & $D_y$ & $D'$  & $C$ & $C_e$ &  $C_{\rev{2}}$ & $C_y$ \\
	\hline
\end{tabular}
\end{table}

\subsection{Behavior at branch point}
\label{sec:branchpoint}
In the vicinity of the branch point, $T=T_{\min}$, 
there exist two solutions, $\alpha_+$ and $\alpha_-$, very close, 
see figure~\ref{fig:ap2ed}.  
Thus they must appear  each other in the variated orbits by eigenfunctions  
as Sim\'{o}'s H obits in $D^H_y$.
Since $\alpha_+$ and $\alpha_-$ are figure-eight choreographic, 
the eigenfunctions have to be $C_e$. 

At $T=T_0=14.4950$ close to $T_{\min}$, 
the solution $\alpha_-$ has slightly lower action 
$S=11.24342$ 
than $S=11.24353$ 
for the solution $\alpha_+$, 
see figure~\ref{fig:ap2ed}.
Thus the solution $\alpha_-$ has eigenfunction $C_e$ with positive eigenvalue 
and inversely
the $\alpha_+$ 
negative. 
%
\begin{figure}
	\centering
	\includegraphics[width=6cm]{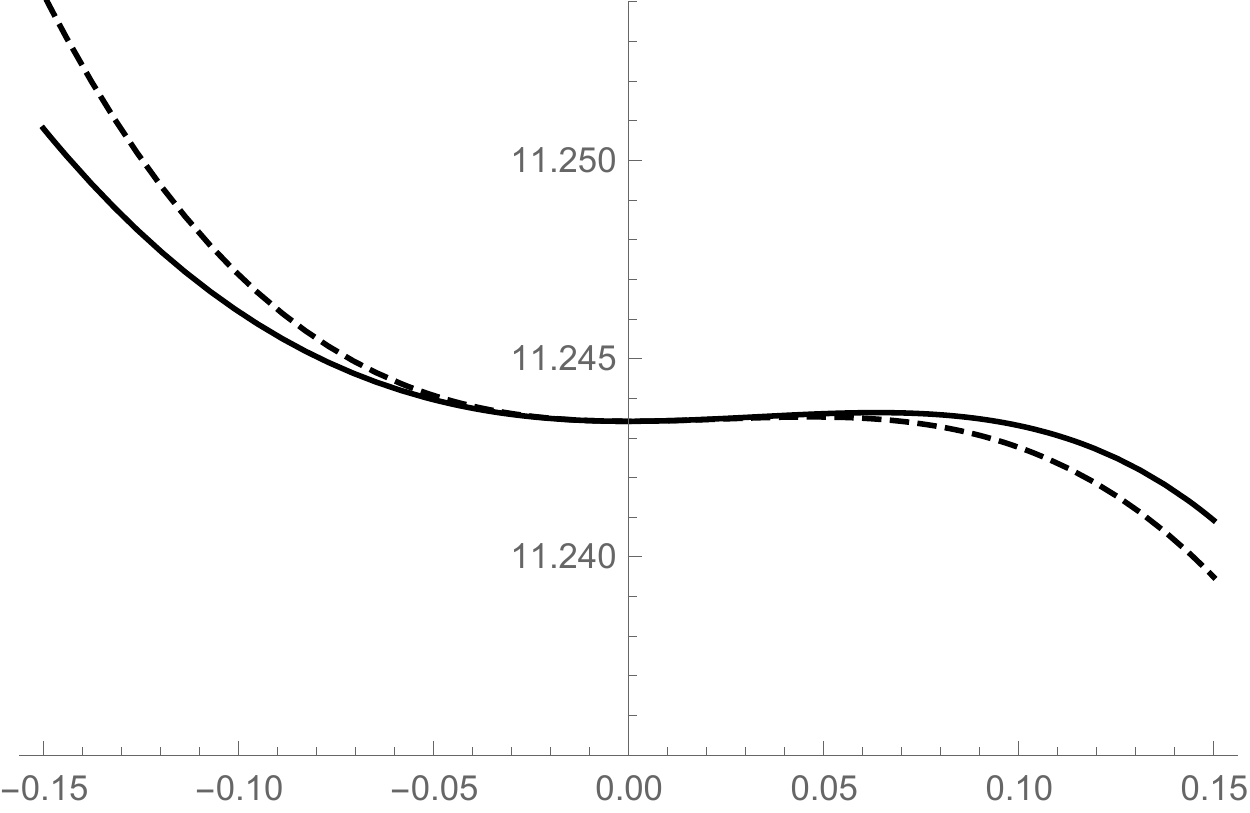} 
	\includegraphics[width=6cm]{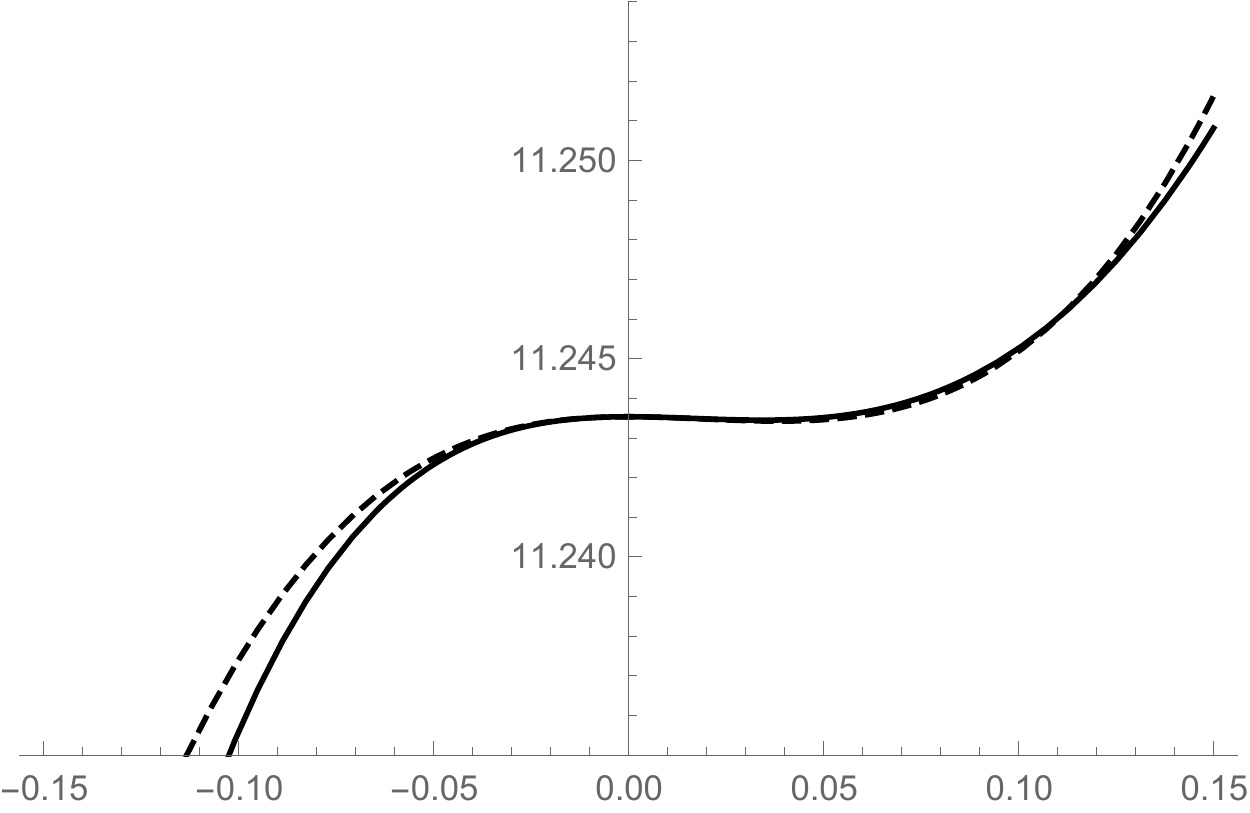} 
	\\
	(a) $q=\alpha_-$, $\lambda=0.30449$ \hspace{1cm} (b) $q=\alpha_+$, $\lambda=-0.448203$
	\caption{
       Solid curve is $S(q+h\psi)$ against $h$ 
       and $\psi$ its eigenfunction of $C_e$ 
       at $T=T_0=14.4950$. 
       (a) There is 
\rev{local minimum} 
        at $h=0$ and 
\rev{maximum} 
       at $h=0.0636$.
       (b) There is 
\rev{local maximum} 
       at $h=0$ and 
\rev{minimum} 
       at $h=0.0343$.
       Dashed curve is $S(q+h\psi)$ truncated at $h^3$ term (\ref{eq:S}).
    }
	\label{fig:ap0682222C6}
\end{figure}
In figure~\ref{fig:ap0682222C6} (a) and (b), 
$S(q+h\psi)$ at $T=T_0$ for $\psi=C_e$ are plotted against $h$
for $q=\alpha_-$ and $\alpha_+$, respectively.
In figure~\ref{fig:ap0682222C6} (a), 
there are 
\rev{local minimum} 
at $h=0$ due to  $\lambda=0.30449>0$
and 
\rev{local maximum} 
at $h=0.0636$, 
and in (b) 
\rev{local maximum} 
due to $\lambda=-0.448203<0$
and
\rev{local minimum} 
at $h=0.0343$.
Both $h$ are close to 
the distance between $\alpha_-$ and $\alpha_+$, 
\begin{equation}\label{eq:h}
  ||q-q'||=0.0428 
\end{equation}
where $q=\alpha_-$, $q'=\alpha_+$, and $||f||=\sqrt{(f,f)}$, 
thus the 
\rev{local minimum} 
and 
\rev{maximum} 
at $h \ne 0$ are considered to be about 
the critical points corresponding to the solutions $\alpha_+$ and $\alpha_-$, respectively.

Dashed curves in figure~\ref{fig:ap0682222C6} are $S(q+h\psi)$ truncated at $h^3$ term
with $S^{(0)}=S(q)$, $S^{(1)}=0$, $S^{(2)}=\lambda$, $S^{(3)}=-2 \lambda / h$ 
where $h$ in $S^{(3)}$ is given by (\ref{eq:posbylambda}).
The positions of 
\rev{local maximum} 
and 
\rev{minimum} 
at $h \ne 0$ of dashed curves are 
calculated by (\ref{eq:posbylambda}) as 
$h=0.0466$ and $0.0384$, respectively. 
Here we have four values of $h$ distributed around (\ref{eq:h}) and  
one of them $h=0.0636$ for the 
\rev{local maximum} 
in figure~\ref{fig:ap0682222C6} (a) is deviated. 
The reason of this distribution is not yet understood.

In figure~\ref{fig:ap072pub}, (f), and in figure~\ref{fig:8all}, (r) are the variated orbit for $C_e$. 
The former may represent $\alpha_-$ if $h$ is suitable, 
however,  at present, the role of the latter for the homogeneous system is unknown.

We note that at the same $T$,
$\alpha_+$ has higher action than $\alpha_-$ by definition,
then as explained above, 
eigenvalue of the $C_e$ in $\alpha_+$ has to be negative 
and $\alpha_-$ positive in the vicinity of $T=T_{\min}$.
In other words, 
following the solution $\alpha_-$ by decreasing $T$, 
its positive eigenvalue of $C_e$ has to change the sign negative at $T=T_{\min}$.
Therefore all $N$, $N_c$ and $N_e$ have to jump by one at $T=T_{\min}$
as shown in figure \ref{fig:alphaIndexTed}.

\subsection{Euler 
	\rev{characteristic}}
\label{sec:euler}
We consider the action manifold $S(f)$ 
in the domain of the figure-eight choreographic function $f$ with period $T$ 
and its Euler 
\rev{characteristic}
\begin{equation}\label{eq:euler}
  \chi_e=\sum_{q'} (-1)^{N_e(q')}
\end{equation} 
where $N_e(q')$ is the Morse index at critical point $q'$ of the manifold,
that is, 
the figure-eight choreographic solution of (\ref{eq:motion}).
According to the theorem by Sbano and Southall \cite{sbano},
there is no figure-eight choreography with $T<T_c$ for some $T_c>0$, thus
there is no critical point 
for $T<T_c$ then $\chi_e=0$.
In the vicinity of $T=T_{\min}$, 
if there is no figure-eight choreography other than $\alpha$
as suggested in \cite{fukuda} 
where $T_c=T_{\min}$ is expected,
we obtain the right hand side
\begin{equation}\label{key}
(-1)^0 + (-1)^1 = 0
\end{equation}
by (\ref{eq:Nep})  and (\ref{eq:Nem}).
This shows the relation (\ref{eq:euler}) holds 
under a common assumption that the Euler 
\rev{characteristic} $\chi_e$ 
is constant 
\rev{for $T>0$}.

\section{Summary and discussions}
\label{sec:summary}
In this paper, 
we solved eigenvalue problem 
(\ref{eq:eigen}) for Morse indices numerically
for the figure-eight choreographies
under homogeneous potential with $a \ge 0$ 
and for the $\alpha$ solution under LJ potential (\ref{eq:LJu}).

The eigenfunctions are classified 
into periodic, choreographic, figure-eight choreographic,
zero and trivial,
and then three kind of Morse indices $N$, $N_c$ and $N_e$ are counted.
We notice that 
the choreographic eigenfunction is non-degenerate and 
the non-choreographic eigenfunction is doubly degenerate.
More detailed analysis of the operator $\hat{H}$ will be published elsewhere \cite{fujiwara}.

We then investigated the correlation of eigenfunctions 
\rev{in $T \ge T_{\min}$} for the solution $\alpha$ under LJ
\rev{and in $a \ge 0$ for homogeneous system}.
For the two eigenfunctions, labeled $D_y^H$ and $C_e$, their variated orbits 
correspond to the real solutions, 
Sim\'{o}'s H and the $\alpha$ solution itself, respectively.

Several questions arise on Sim\'{o}'s H orbit.
How does Sim\'{o}'s H orbit change when $a$ is varied?
Does periodic orbit corresponding to Sim\'{o}'s H orbit exist for LJ system?
Is there non-choreographic orbit 
corresponding to the $D$ with $\lambda=-0.0116029$ 
in figure~\ref{fig:8all} (a) and (b) for $a=1$,
say, a non-symmetric H orbit since $D$ is not symmetric in $y$-axis?
How does it behave at $a=a_1$ where the eigenvalue changes to positive? 

\begin{figure}
	\centering
	\includegraphics[width=6cm]{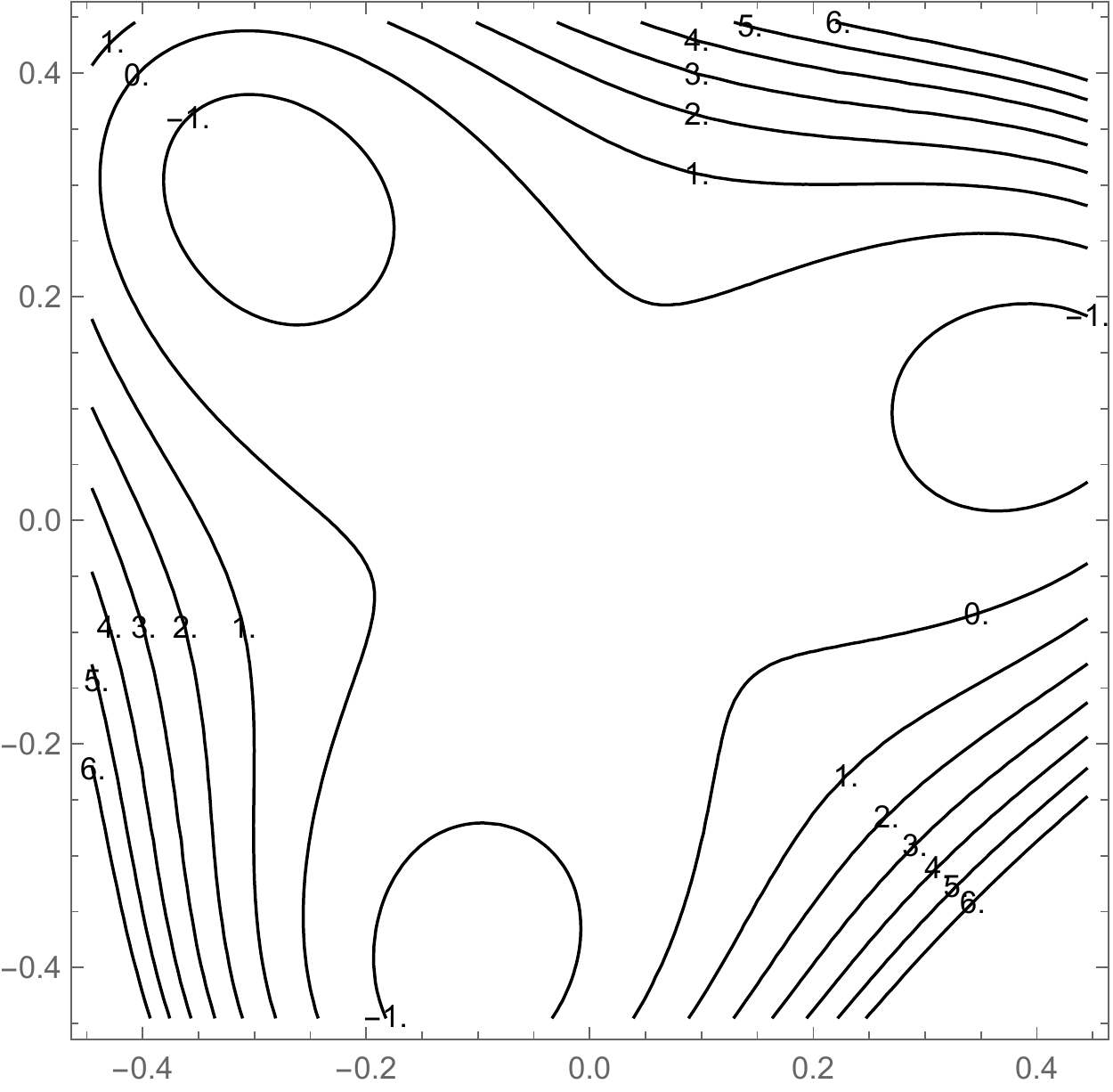} 
	\caption{
		Contour plot for action $S(q+h \psi^{(\Theta)})$.  
		$q$ is the figure-eight choreography under homogeneous potential with $a=a_-=0.994$ and 
		the eigenfunction $\psi^{(\Theta)}$ at $\rev{s}=3$ for $D_y^H$.
		Horizontal and vertical axis are $h \cos\Theta$ and $h \sin\Theta$, respectively.
		Contours are labeled by $(S(q+h \psi^{(\Theta)})-S(q)) \times 10^5$.
	}
	\label{fig:cont0994}
\end{figure}
At $a=a_0$, the eigenvalue of the $D_y^H$ correlated to the Sim\'{o}'s H orbit 
changes the sign 
as (\ref{eq:N(a)}).
%
In figure~\ref{fig:cont0994}, 
contour plot of the action $S(q+h \psi^{(\Theta)})$
for $\psi^{(\Theta)}=D_y^H$ at $a=a_-=0.994<a_0$ 
where eigenvalue for $D_y^H$ is negative 
are shown. 
The three fold symmetry and 
\rev{local minima} 
around origin in contour is observed, 
suggesting the existence of Sim\'{o}'s H orbit at $a=a_-$.

On the other hand, 
if there exists no Sim\'{o}'s H at $a=a_{-}$,  
the Euler 
\rev{characteristic} of the action manifold in the domain of the periodic function 
\begin{equation}\label{key}
  \chi=\sum_{q'} (-1)^{N(q')}
\end{equation}
at $a=a_{-}$ is  $\chi=(-1)^{N}=(-1)^4=1$ by (\ref{eq:N(a)}).
However, at $a=1>a_0$, 
$\chi$ is even since $N=2$ by (\ref{eq:N(a)}) 
and 
there are three critical points $q'$'s for Sim\'{o}'s H orbits, 
\rev{$(-1)^2+3(-1)^{N(q')}=1\pm3$}.
Thus the conservation of the Euler 
\rev{characteristic} $\chi$ 
around $a=a_0$ also supports the existence of the Sim\'{o}'s H at $a=a_-$.
Then we expect that the Sim\'{o}'s H orbit will exist in the both sides of $a=a_0$.

For the solution $\alpha$ under LJ potential, 
we found that Morse index $N_e$ for figure-eight choreography 
is 0 for $\alpha_-$ and 1 for $\alpha_+$ and that it changes at minimum of $T$.
This behavior of $N_e$ is consistent with 
the Euler 
\rev{characteristic} $\chi_e=0$ of the action manifold  given by 
the theorem by Sbano and Southall \cite{sbano}.
We expect from this results
that  the calculation of Morse indices 
for the other series of solutions, $\beta$, $\gamma$,
$\delta$, ... found in \cite{fukuda} 
helps to understand their structures and relations through action manifold.
\del{Although we did not discuss the results (\ref{eq:N})
and (\ref{eq:Nc}) on $N$ and $N_c$,
related to the non-choreographic solutions and 
choreographic solution in general symmetry, 
here, but they are also interesting.}
%
\rev{
	Although we investigated what happened at zero of the eigenvalue 
	for the eigenfunction $C_e$ under constant Euler characteristic $\chi_e$ 
	in section \ref{sec:branchpoint} and \ref{sec:euler}, 
	we did not for the other eigenfunctions yet: 
	choreographic ones, $C$, $C_y$ and $C_2$ 
	under $\chi_c=\sum_{q'}(-1)^{N_c(q')}$,
	and non-choreographic ones, $D_y^H$, $D_y$, $D$ and $D'$ 
	under $\chi=\sum_{q'}(-1)^{N(q')}$.
	They are also interesting and more studies will be needed in future.}

\rev{
The analysis in this paper was performed with a fixed value of the strength of the potential terms, 
(\ref{eq:homou}), the repulsive term in the LJ potential  (\ref{eq:LJu}) and the attractive. 
However, since the changes of the strength of potential  terms are 
identical to the scale transformation in time and length, 
the analysis is not changed if the strength is varied. }

\section*{Acknowledgments}
The research of HF was supported by
Grant-in-Aid for Scientific Research 17K05146 JSPS.

\end{document}